%% file: paper.tex
\documentclass[10pt,conference,final]{IEEEtran}
\IEEEoverridecommandlockouts

\usepackage{cite}
\usepackage{amsmath,amssymb,amsfonts}
\usepackage{algorithmic}
\usepackage{graphicx}
\usepackage{textcomp}
\usepackage{xcolor}
\def\BibTeX{{\rm B\kern-.05em{\sc i\kern-.025em b}\kern-.08em
    T\kern-.1667em\lower.7ex\hbox{E}\kern-.125emX}}

\iftrue
\usepackage{hyperref}

\hypersetup{
  colorlinks   = true, 
  urlcolor     = blue, 
  linkcolor    = black, 
  citecolor   = black 
}
\fi

\usepackage[utf8]{inputenc}
\usepackage{cleveref}
\crefname{section}{§}{§§}
\Crefname{section}{§}{§§}

\usepackage{pifont}
\usepackage{setspace}
\usepackage{algorithmic}
\usepackage{algorithm2e} 

\usepackage{bold-extra}

\usepackage{balance}

\usepackage{adjustbox,multirow}

\newcommand{\nextnr}{\stepcounter{AlgoLine}\ShowLn}
\makeatletter
\newcommand{\removelatexerror}{\let\@latex@error\@gobble}
\makeatother

\newlength{\textfloatsepsave}
\newlength{\floatsepsave}
\usepackage{xcolor} 
\usepackage{tikz}
\usetikzlibrary{fit,calc}
\newcommand*{\tikzmk}[1]{\tikz[remember picture,overlay,] \node (#1) {};\ignorespaces}
\newcommand{\boxit}[1]{\tikz[remember picture,overlay]{\node[yshift=1pt,fill=#1,opacity=.20,fit={(A)($(B)+(.80\linewidth,0.9\baselineskip)$)}] {};}\ignorespaces}
\colorlet{pink}{red!40}
\colorlet{blue4}{cyan!30}
\colorlet{blue2}{cyan!60}
\colorlet{blue3}{cyan!90}

\colorlet{gray1}{gray!45}
\colorlet{gray2}{gray!90}

\usepackage{framed}
\definecolor{shadecolor}{gray}{0.9}
\newcommand{\boxedspit}[1]{
\vspace{-2mm}
\begin{shaded*}
\vspace{-2mm}
#1 
\vspace{-2mm}
\end{shaded*}
\vspace{-2mm}
}

\begin{document}

\newcommand{\fuzzysat}{{\sc Fuzzy-Sat}}
\newcommand{\fuzzolic}{{\sc Fuzzolic}}
\newcommand{\qsym}{{\sc Qsym}}
\newcommand{\redqueen}{{\sc RedQueen}}
\newcommand{\angora}{{\sc Angora}}
\newcommand{\afl}{{\sc AFL}}
\newcommand{\aflpp}{{\sc AFL++}}
\newcommand{\jfs}{{\sc JFS}}
\newcommand{\angr}{{\sc Angr}}
\newcommand{\binsec}{{\sc BinSec}}
\newcommand{\fuzzball}{{\sc FuzzBall}}
\newcommand{\klee}{{\sc Klee}}
\newcommand{\stwoe}{{\sc S2E}}
\newcommand{\symcc}{{\sc SymCC}}
\newcommand{\slf}{{\sc SLF}}
\newcommand{\eclipser}{{\sc Eclipser}}

\def\rem#1{}
\newcommand\mdoubleplus{\mathbin{+\mkern-10mu+}}

\newcommand{\edits}[1]{#1}

\title{
Fuzzing Symbolic Expressions\\ 
\thanks{This paper is supported in part by European Union's Horizon 2020 research and innovation 
programme (grant agreement No. 830892, project SPARTA).}
}

\author{
\IEEEauthorblockN{Luca Borzacchiello}
\IEEEauthorblockA{\textit{DIAG Department} \\
\textit{Sapienza University of Rome}\\
Rome, Italy \\
borzacchiello@diag.uniroma1.it}
\and
\IEEEauthorblockN{Emilio Coppa}
\IEEEauthorblockA{\textit{DIAG Department} \\
\textit{Sapienza University of Rome}\\
Rome, Italy \\
coppa@diag.uniroma1.it}
\and
\IEEEauthorblockN{Camil Demetrescu}
\IEEEauthorblockA{\textit{DIAG Department} \\
\textit{Sapienza University of Rome}\\
Rome, Italy \\
demetres@diag.uniroma1.it}
}

\maketitle

\begin{abstract}
Recent years have witnessed a wide array of results in software testing, exploring different approaches and methodologies ranging from fuzzers to symbolic engines, with a full spectrum of instances in between such as concolic execution and hybrid fuzzing. A key ingredient of many of these tools is Satisfiability Modulo Theories (SMT) solvers, which are used to reason over symbolic expressions collected during the analysis. In this paper, we investigate whether techniques borrowed from the fuzzing domain can be applied to check whether symbolic formulas are satisfiable in the context of concolic and hybrid fuzzing engines, providing a viable alternative to classic SMT solving techniques. We devise a new approximate solver, \fuzzysat, and show that it is both competitive with and complementary to state-of-the-art solvers such as Z3 with respect to handling queries generated by hybrid fuzzers.
\end{abstract}


\input{intro}
\input{related}

\input{approach}
\input{implem}

\input{eval}

\input{conclusions}

\bibliographystyle{IEEEtran}
\bibliography{IEEEabrv,mybibfile}
\balance

\end{document}

%% file: intro.tex
\section{Introduction}
\label{se:intro}

The automatic analysis of modern software, seeking for high coverage and bug detection is a complex endeavor. 
Two popular approaches have been widely explored in the literature: on one end of the spectrum, {\em coverage-guided fuzzing} starts from an input seed and applies simple transformations (mutations) to the input, re-executing the program to be analyzed to increase the portion of explored code. The approach works particularly well when the process is guided and informed by code coverage, with a nearly-native execution time per explored path~\cite{AFL,AFLPP}. On the other end of the spectrum, {\em symbolic execution} (SE) assigns symbolic values to input bytes and builds expressions that describe how the program manipulates them, resorting to satisfiability modulo theories (SMT)~\cite{Barrett2018} solver queries to reason over the program, e.g., looking for bug conditions. A popular variant of SE is {\em concolic execution} (CE), which concretely runs one path at a time akin to a fuzzer, collecting branch conditions along the way~\cite{QSYM, SYMCC}. By systematically negating these conditions, it steers the analysis to take different paths, aiming to increase code coverage. The time per executed path is 
higher than fuzzing but the aid of a solver allows for a smaller number of runs. 


Different ideas have been proposed to improve the effectiveness of analysis tools by combining ideas from both fuzzing and SE somewhere in the middle of the spectrum. As a prominent example, {\em hybrid fuzzing} couples a fuzzer with a symbolic executor to enable the exploration of complex branches~\cite{stephens2016driller, QSYM}. Compared to base fuzzing, this idea adds a heavy burden due to the lack of scalability of symbolic execution. It is therefore of paramount importance to speed up the symbolic part of the exploration.



While there is no clear winner in the software testing 
spectrum, 
tools that hinge upon an SMT solver get a high price to pay in terms of running time, limiting their throughput.

\medskip
\noindent{\bf Contributions.} As a main contribution, this paper addresses the following research question:


\begin{quote} {\em can we avoid using accurate but costly SMT solvers, replacing them with an approximate solver to test satisfiability in the context of software testing?}
\end{quote}

\noindent We attempt to positively answer this question, devising \fuzzysat, an approximate solver that borrows ideas from the fuzzing domain. Our solver is tailored to the symbolic expressions generated by concolic engines and can replace classic SMT solvers in this context. By analyzing the expressions contained in symbolic queries, \fuzzysat\ performs {\em informed} mutations to possibly generate new valuable inputs.
To demonstrate the potential behind \fuzzysat, we present \fuzzolic, a new hybrid fuzzer based on QEMU. To show that \fuzzysat\ can be used in other frameworks, we integrate it also in QSYM~\cite{QSYM}.
In our experimental evaluation:


\begin{enumerate}
\item we compare \fuzzysat\ to the SMT solver Z3~\cite{Z3-TACS08} and the approximate solver \jfs~\cite{JFS} on queries issued by QSYM, which we use as a mature baseline. Our results suggest that \fuzzysat\ can provide a nice tradeoff between speed and solving effectiveness, i.e., the number of queries found satisfiable by a solver.


\item we show that \fuzzysat\ allows \qsym\ to find more bugs on the LAVA-M dataset~\cite{LAVAM} compared to Z3.



\item we evaluate \fuzzolic\ on 12 real-world programs against state-of-the-art fuzzers including \aflpp~\cite{AFLPP}, \eclipser~\cite{ECLIPSER-ICSE19}, and \qsym, showing that it can reach higher code coverage than the competitors. 
%
\end{enumerate}

To facilitate extensions of our approach, we make our contributions available at:
\begin{center}
{\tt \url{https://season-lab.github.io/fuzzolic/}}
\end{center}

%% file: related.tex
\section{Background}
\label{se:related}

\fuzzysat\ takes inspiration from
two popular software testing techniques~\cite{art-software-testing}: symbolic execution~\cite{SurveySymbolic} and coverage-based grey-box fuzzing~\cite{fuzzing-book}. We now review the inner-workings of these two approaches, focusing on recent works that are tightly related to the ideas explored in this paper.

{\bf Symbolic execution.} The key idea behind this technique is to execute a program over {\em symbolic}, rather than {\em concrete}, inputs. Each symbolic input can, for instance, represent a byte read by the program under test from an input file and initially evaluate to any value admissible for its data type (e.g., $[0, 255]$ for an unsigned byte). SE builds expressions to describe how the program manipulates the symbolic inputs, resorting to SMT solver queries to reason over the program state. In particular, when a branch condition $b$ is met during the exploration, SE  checks using the solver whether both directions can be taken by the program for some values of the inputs, forking the execution state in case of a positive answer. When forking, SE updates the list of {\em path constraints} $\pi$ that must hold true in each state: $b$ is added in the state for the {\em true} branch, while $\neg b$ is added to the state for the {\em false} branch. At any time, the symbolic executor can generate concrete inputs, able to reproduce the program execution related to one state, by asking the solver an assignment for the inputs given $\pi$.

SE can be performed on binary code (e.g., \angr~\cite{ANGR-SSP16}, S$^{2}$E~\cite{S2E-TOCS12}) or on more high-level representations of a program (e.g., LLVM IR in \klee~\cite{KLEE-OSDI08}, Java bytecode in SPF~\cite{SPF}). 
\edits{Besides software testing, SE has been extensively used during the last decade in the context of cybersecurity~\cite{MAYHEM,BCDD-CSCML17,BCDD-CSCML19}.}

{\bf Concolic execution.} A twist of SE designed with scalability in mind
is concolic execution~\cite{SAGE-NDSS08}, 
%
which given a concrete input $i$, analyzes symbolically only the execution path taken by the program when running over $i$. To generate new inputs, the concolic executor can query an SMT solver using $\neg b \wedge \pi$, where $b$ is a branch condition taken by the program in the current path while $\pi$ is the set of constraints from the branches previously met along the path. 
A benefit from this approach is that the concolic executor only needs to query the solver for one of the two branch directions, as the other one is taken by the path under analysis. Additionally, if the program is actually executed concretely in parallel during the analysis, the concolic engine can at any time trade accuracy for scalability, by concretizing some of the input bytes and make progress in the execution using the concrete state. For instance, 
 when analyzing a complex library function, the concolic engine may {\em concretize} the arguments for the function and execute it concretely, without issuing any query or making $\pi$ more complex due to the library code but possibly giving up on some alternative inputs due to the performed concretizations.


A downside of most concolic executors is that they restart from scratch for each input driving the exploration, thus
repeating analysis work across different runs. 
To mitigate this problem, \qsym~\cite{QSYM} has proposed a concolic executor built through dynamic binary instrumentation (DBI) that cuts down the time spent for running the program by maintaining only the symbolic state and offloads completely the concrete state to the native CPU. Additionally, it simplifies the symbolic state 
by concretizing symbolic addresses~\cite{CDD-ASE17,BCE-STVR19} but also generates inputs that can lead the program to access alternative memory locations. More recently, \symcc~\cite{SYMCC} has improved the design of \qsym\ by proposing a source-based instrumentation approach that further reduces the emulation time.


\smallskip
{\bf Approximate constraint solving.} 
Many queries generated by concolic executors are either unsatisfiable or cannot be solved within a limited amount of time~\cite{QSYM}. This often is due to the complex constraints contained in $\pi$, which can impact the reasoning time even when the negated branch condition is quite simple. For this reason, \qsym\ has introduced {\em optimistic solving} that, in case of failure over $\neg b \wedge \pi$ due to unsatness or solving timeout, submits to the solver an additional query containing only $\neg b$: by patching the input $i$ (used to drive the exploration) in a way that makes $\neg b$ satisfied, the executor is often able to generate valuable inputs for a program.



A different direction is instead taken by \jfs~\cite{JFS}, which builds on the experimental observation that SMT solvers can struggle on queries that involve floating-point values. \jfs\ thus proposes to turn the query into a program, which is then analyzed using coverage-based grey-box fuzzing. More precisely, the constructed program has a point that is reachable if and only if the program's input satisfies the original query. The authors show that \jfs\ is quite effective on symbolic expressions involving floating-point values but it struggles on integer values when compared to traditional SMT solvers.

Two very recent works, {\sc Pangolin}~\cite{PANGOLIN-SP20} and {\sc Trident}~\cite{TRIDENT-ISSTA20}, devise techniques to reduce the solving time in CE. {\sc Pangolin} transforms constraints into a {\em polyhedral path abstraction}, modeling the solution space as a polyhedron and using, e.g., sampling to find assignments. {\sc Trident} instead exploits interval analysis to reduce the solution space in the SMT solver. Their implementations have not been released yet.

\smallskip
{\bf Coverage-based grey-box fuzzing.} An orthogonal approach to SE is coverage-based grey-box fuzzing (CGF). Given an input queue $q$ (initialized with some input seeds) and a program $p$, CGF picks an input $i$ from $q$, randomly mutates some of its bytes to generate $i'$ and then runs $p$ over $i'$: if new code is executed (covered) by $p$ compared to previous runs on other inputs, then CGF deems the input {\em interesting} and adds it to $q$. This process is then repeated endlessly, looking for crashes and other errors during the program executions.


American Fuzzy Lop (\afl)~\cite{AFL} is the most prominent example of CGF. 
To track the coverage, it can dynamically instrument at runtime a binary or add source-based instrumentation at compilation time. The fuzzing process for each input is split into two main stages. In the first one, \afl\ scans the input and {\em deterministically} applies for each position a set of mutations, testing the effect of each mutation on the program execution in isolation. In the second stage, 
\afl\ instead 
performs several mutations in sequence, i.e., {\em stacking} them, over the input, {\em non deterministically} choosing which mutations to apply and at which positions. The mutations in the two stages involve simple and fast to apply transformations such as flipping bits, adding or subtracting constants, removing  bytes, combining different inputs, and several others~\cite{AFL}. 


\smallskip
{\bf Hybrid fuzzing.} Although CGF fuzzers have found thousands of bugs in the last years~\cite{aflsmart-tse,google-oss-fuzz}, there are still scenarios where  their mutation strategy is not effective. For instance, they may struggle on checks against magic numbers, whose value is unlikely to be generated with random mutations. As these checks may appear early in the 
execution, fuzzers may soon {\em get stuck} and stop producing interesting inputs. For this reason, a few works have explored combinations of fuzzing with symbolic execution, proposing {\em hybrid fuzzing}. {\sc Driller}~\cite{DRILLER-NDSS16} alternates \afl\ and \angr, temporarily switching to the latter when the former is unable to generate new interesting inputs for a specific budget of time. 
\qsym\ proposes instead to run a concolic executor in parallel with \afl, allowing the two components to share their input queues and continuously benefit from the work done by each other.  


\smallskip
{\bf Recent improvements in coverage-guided fuzzing.} During the last years, a large body of works has extended CGF, trying to make it more effective without resorting to heavyweight analyses such as symbolic execution. {\sc Laf-Intel}~\cite{lafintel} splits multi-byte checks into single-byte comparisons, helping the fuzzer track the intermediate progress when reasoning on a branch condition. {\sc Vuzzer}~\cite{vuzzer} integrates dynamic taint analysis (DTA)~\cite{debray-scam14} into the fuzzer to identify which bytes influence the operands in a branch condition, allowing it to bypass, e.g., checks on magic numbers. \angora~\cite{angora} further improves this idea by performing multi-byte DTA and using gradient descent to effectively mutate the tainted input bytes.



As DTA can still put a high burden on the fuzzing strategy, some works have recently explored lightweight approximate analyses that can replace it. \redqueen~\cite{redqueen} introduces the concept of {\em input-to-state correspondence}, which captures the idea that input bytes often flow directly, or after a few simple encodings (e.g., byte swapping), into comparison operands during the program execution. To detect this kind of input dependency, \redqueen\ uses {\em colorization} that inserts random bytes into the input and then checks whether some of these bytes appear, as is or after few simple transformations, in the comparison operands when running the program. 
Input-to-state relations can be exploited to devise effective mutations and bypass several kinds of validation checks.


{\sc Weizz}~\cite{WEIZZ} explores instead a different approach that flips one bit at a time on the entire input, checking after each bit flip which comparison operands have changed during the program execution, possibly suggesting a dependency between the altered bit and the affected branch conditions. While more accurate than colorization, this approach may incur a large overhead, especially in presence of large inputs. 
Nonetheless, {\sc Weizz} is willing to pay this price as the technique allows it to also heuristically locate fields and chunks within an input, supporting {\em smart mutations}~\cite{aflsmart-tse} to effectively fuzz applications processing structured input formats.


\slf~\cite{SLF} exploits a bit flipping strategy similar to {\sc Weizz} to generate valid inputs for an application even when no meaningful seeds are initially available for it. Thanks to the input dependency analysis, \slf\ can identify fields into the input and then resort to a gradient-based multi-goal search heuristic to deal with interdependent checks in the program.


\eclipser~\cite{ECLIPSER-ICSE19} identifies a dependency between an input byte $i_k$ and a branch condition $b$ whenever the program decision on $b$ is affected when running the program on inputs containing different values for $i_k$. \eclipser\ builds approximate path constraints by modeling each branch condition met along the program execution as an interval. In particular, given a branch $b$, it generates a new input using a strategy similar to concolic execution, by looking for input values that satisfy the interval from $\neg b$ as well as any other interval from previous branches met along a path. 
To find input assignments, \eclipser\ does not use an SMT solver but resorts to lightweight techniques that work well in presence of intervals generated by linear or monotonic functions.


%% file: approach.tex
\section{Approach}
\label{se:approach}




Recent coverage-guided fuzzers perform input mutations based of a knowledge on the program behavior that goes beyond the simple code coverage. Concolic executors by design build an accurate description of the program behavior, i.e., symbolic expressions, but outsource completely the reasoning to a powerful but expensive SMT solver, which is typically treated as a black box. In this paper, we explore the idea that a concolic executor can learn from the symbolic expressions that it has built and use the acquired knowledge to apply simple but fast input transformations, possibly solving queries without resorting to an SMT solver. The key insight is that given a query $\neg b \wedge \pi$, the input $i$ that has driven the concolic exploration satisfies by design $\pi$. Hence, we propose to build using input mutations a new test case $i'$ that satisfies $\neg b$ and is similar enough to $i$ so that $\pi$ remains satisfied by $i'$.  In the remainder of this section, we present the design of \fuzzysat, an approximate solver that explores this direction by borrowing ideas from the fuzzing domain to efficiently solve queries generated by concolic execution. 



\subsection{Reasoning primitives for concolic execution}
\label{ss:primitives}




While SMT solvers typically offer a rich set of solving primitives, enabling reasoning on formulas generated from quite different application contexts, concolic executors such as \qsym\ are instead built on top of a few but essential primitives. In this paper, we focus on these primitives without claiming that \fuzzysat\ can replace a full-fledged SMT solver in a general context. \fuzzysat\ exposes the following primitives:

\begin{itemize}

\item {\sc Solve}$(e, \pi, i, opt)$: returns an assignment for the symbolic inputs in $e\wedge\pi$ such that the expression $e\wedge\pi$ is satisfiable. 
The flag $opt$ indicates whether optimistic solving should be performed in case of failure. 
This primitive is used by concolic engines when negating a branch condition $b$, hence $e=\neg b$.

\item {\sc SolveMax}$(e, \pi, i)$ (resp. {\sc SolveMin}$(e, \pi, i)$): returns an assignment that maximizes (resp. minimizes) $e$ while making $\pi$ satisfiable. Concolic executors use these primitives before concretizing a symbolic memory address $e$ to keep the exploration scalable. These functions are thus used to generate alternative inputs that steer the program to read/write at boundary addresses.



\item {\sc SolveAll}$(e, \pi, i)$: combines {\sc SolveMin} and {\sc SolveMax}, yielding intermediate assignments identified during the reasoning process as well. This primitive is valuable in the presence of symbolic memory addresses accessing a jump table or when the instruction pointer becomes symbolic during the exploration.

\end{itemize}

Two main aspects differentiate these primitives in \fuzzysat\ with respect to their counterpart from an SMT solver. 

First, \fuzzysat\ is an approximate solver and thus it cannot guarantee that no valid assignment exists in case of failure of {\sc Solve}, i.e., \fuzzysat\ cannot prove that an expression $e\wedge\pi$ is unsatisfiable. Similarly, given an expression $e$, \fuzzysat\ may fail to find its global minimum/maximum value or to enumerate assignments for all its possible values.

Another crucial difference is that \fuzzysat\ requires that the concolic engine provides the input test case $i$ that was used to steer the symbolic exploration of the program under test. This is essential as \fuzzysat\ builds assignments by mutating the test case $i$ based on facts that are learned when analyzing $e$ and $\pi$. Given an assignment $a$ returned by \fuzzysat, a new input test case $i'$ can be built by patching the bytes in $i$ that are assigned by $a$.


\subsection{Overview}
\label{ss:overview}

\begin{figure}[t]
\centerline{\includegraphics[height=2.7cm]{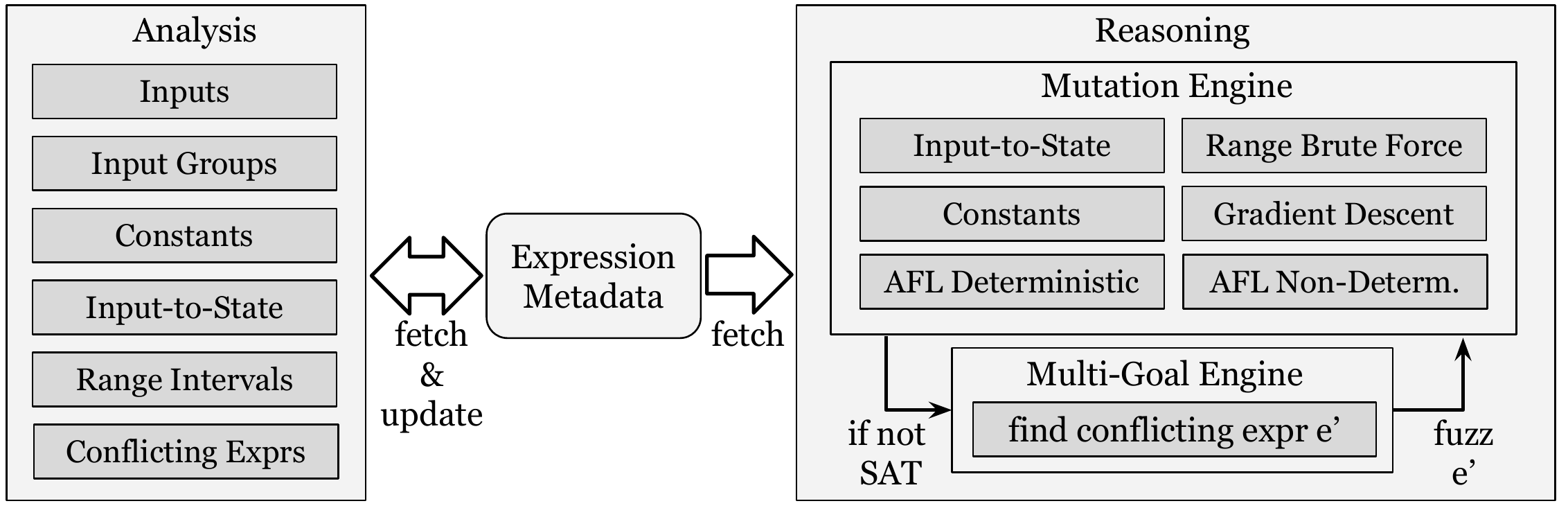}}\vspace{-2mm}
\caption{Internal architecture of \fuzzysat.\label{fig:structure}}
\end{figure}

{\bf Architecture.} To support the primitives presented in Section~\ref{ss:primitives}, the architecture of \fuzzysat\ (Figure~\ref{fig:structure}) has been structured around three main building blocks: the {\em analysis stage}, the {\em expression metadata}, and the {\em reasoning stage}.

The analysis stage (\cref{ss:analysis}) is designed to analyze symbolic expressions, extracting valuable knowledge to use during the reasoning stage. It starts by identifying which input bytes $i_k$ from the input $i$ are involved in an expression and how they are grouped.
It detects input-to-state relations (\cref{se:related}) and collects constants appearing in the expression for later use in the mutation phase. Expressions that constrain the interval of admissible values for a set of inputs, dubbed {\em range constraints}, such as $i_0 < 10$, are identified to keep track of the {\em range intervals} over the symbolic inputs. Finally, this stage detects whether the current expression shares input bytes with other expressions previously processed by the analysis component, possibly pinpointing {\em conflicts} that may result when mutating these bytes during the reasoning stage.

The expression metadata maintains the knowledge of \fuzzysat\ on the expressions processed by the analysis stage over time. Internally, it is implemented as a set of data structures optimized for fast lookup of different kinds of properties related to an expression (and its subexpressions).
It is updated by the analysis stage and queried by both stages. 

Finally, the reasoning stage is where \fuzzysat\ exploits the knowledge over the expressions to effectively fuzz the input test case and possibly generate valid assignments. To reach this goal, a mutation engine (\cref{ss:reasoning}) is used to perform a set of transformations over the input bytes involved in an expression $e$ looking for an assignment that satisfies $e$ and $\pi$ ({\sc Solve}) or maximizes/minimizes $e$ while satisfying $\pi$ (other primitives). When this step finds assignments for $e$, but none of them satisfies $\pi$, then \fuzzysat\ performs a multi-goal strategy, which is not limited to changing the input bytes involved in $e$, but attempts to alter other input bytes that are involved in conflicting expressions present in $\pi$. 

\input{algorithms/solve_one}


{\bf Implementing the reasoning primitives.}
Algorithm~\ref{alg:solve-one} shows the interplay of these three components in \fuzzysat\ when considering the primitive {\sc Solve}. Lines 1 and 2 execute the analysis stage by invoking the {\sc Analyze} function on $\pi$ and $e$, respectively. {\sc Analyze} updates the expression metadata M, adding any information that could be valuable during the reasoning stage. Since concolic engines would typically call {\sc Solve} several times during the symbolic exploration, providing each time a $\pi$ that is the conjunction of branch conditions met along the path and which have been already analyzed by \fuzzysat\ in previous runs of {\sc Solve},
the call at line 1 does not lead \fuzzysat\ to perform any work in most scenarios as the expression metadata M already has a cache containing knowledge about expressions in $\pi$.

Lines 3-18 instead comprise the reasoning stage and can be divided into three main phases. First, the {\sc Mutate} function is called at line 3 to run the mutation engine, restricting the transformations on input bytes that are involved in the expression $e$. When {\sc Mutate} finds an assignment $a$ that satisfies both $e$ and $\pi$, {\sc Solve} returns it at line 4 without any further work. On the other hand, when $a$ is invalid but some assignments SA found by {\sc Mutate} make at least $e$ satisfiable, then {\sc Solve} starts the multi-goal phase (lines 5-15). To this end, \fuzzysat\ uses function {\sc PickBestAssignment} to select the best candidate assignment $a$  from SA
\footnote{We pick an $a$ that maximizes the number of expressions satisfied in $\pi$.}
 and then {\em fixes} the input bytes 
assigned by $a$ using function {\sc FixInputBytes} to prevent further calls of {\sc Mutate} from altering these bytes. It then reruns the mutation engine considering an expression $e'$ which has been marked as in conflict with $e$ during the analysis stage. This process is repeated as long as three conditions hold: (a) $e \wedge \pi$ is not satisfied (line 11), (b) {\sc Mutate} returns at least one assignment in SA for $e'$ (line 13), and (c) there is still a conflicting expression left to consider (condition at line 9). 

The multi-goal strategy in \fuzzysat\ employs a greedy approach without ever performing backtrack (e.g., reverting the effects of {\sc FixInputBytes} in case of failure) as it trades accuracy for scalability. Indeed, 
\fuzzysat\ builds on the intuition that by altering a {\em few} bytes from the input test case $i$
, it is possible in several cases to generate valid assignments. Additionally, 
since many queries generated by a concolic engine are unsatisfiable, increasing the complexity of this strategy would impose a large burden on \fuzzysat.

The last phase of {\sc Solve} (lines 16-18) has been devised to support optimistic solving in \fuzzysat. When the Boolean $opt$ is true, \fuzzysat\ returns the last candidate assignment found by the mutation engine, which by design satisfies the expression $e$. However, since the previous calls to the mutation engine in {\sc Solve} may have failed to find an assignment $a$ for $e$ due to the constraints resulting from the analysis of expressions from $\pi$, \fuzzysat\ as last resort uses a variant of the function {\sc Mutate}, called {\sc MutateOpt}, that ignores these constraints and exploits only knowledge resulting from $e$ when performing transformations over the input bytes.

The other reasoning primitives ({\sc SolveMin}, {\sc SolveMax}, and {\sc SolveAll}, respectively) follow a workflow similar to {\sc Solve} and we do not present their pseudocode due to lack of space. In the remainder of this section, we focus on the internal details of functions {\sc Analyze} (\cref{ss:analysis}) and {\sc Mutate} (\cref{ss:reasoning}), which are crucial core elements of \fuzzysat.

\input{approach-analysis}
\input{approach-reasoning}

\subsection{Discussion}

Similarly to fuzzers using dynamic taint analysis, 
\fuzzysat\ restricts mutations over the bytes that affect branch conditions during the program execution. However, it does not only understand {\em which} bytes influence the branch conditions but also reasons on {\em how} they affect them, possibly devising more effective mutations. 

\fuzzysat\ shares traits with \angora, \slf, and \eclipser\ by integrating mutations based on gradient descent, a multi-goal strategy, and range intervals, respectively. Nevertheless, these techniques have been revisited and refined to work over symbolic constraints, which accurately describe the program state and are not available to these fuzzers.

\fuzzysat\ exposes primitives that are needed by concolic executors and that are typically offered by SMT solvers but it implements them in a fundamentally different way inspired by fuzzing techniques, trading accuracy for scalability.

Finally, \fuzzysat\ shares the same spirit of \jfs\ but takes a rather different approach. While \jfs\ builds a bridge between symbolic execution and fuzzers by turning expressions into a program to fuzz, \fuzzysat\ is designed to merge these two worlds, possibly devising {\em informed} mutations that are driven by
the knowledge 
acquired by analyzing the expressions.

%% file: algorithms/solve_one.tex

\begin{figure}[t]
\vspace{-2mm}
\removelatexerror
\begin{algorithm}[H]
 \footnotesize
 \DontPrintSemicolon
 \SetAlgoNoEnd
 \SetAlgoNoLine
 \SetNlSkip{-0.5em}
 \SetKwFunction{FMain}{\sc Solve}
 \SetKwProg{Fn}{function}{:}{}
 \Fn{\FMain{e, $\pi$, i, opt}}{   
 \tikzmk{A} 
 \nextnr M $\gets$ {\sc Analyze}($\pi$, M)\;
 \nextnr M $\gets$ {\sc Analyze}(e, M)\; 
 \tikzmk{B}
 \boxit{gray1}
 \tikzmk{A} 
 \nextnr a, SA $\gets$ {\sc Mutate}(e, $\pi$, i, M)\;
 \nextnr \lIf{{\upshape a} {\upshape is not} {\upshape NULL}}{\Return{a}}
 \tikzmk{B}
 \boxit{gray2}
 \tikzmk{A} 
 \nextnr a $\gets$ {\sc PickBestAssignment}($\pi$, SA)\;
 \nextnr \If{{\upshape a} {\upshape is not} {\upshape NULL}}{\Indmm
            \nextnr M$'$ $\gets$ {\sc FixInputBytes}(a, M)\;
            \nextnr CC $\gets$ {\sc GetConflictingExpressions}(e, $\pi$, M$'$)\;
            \nextnr \For{{\upshape e$'$}$\,\in\,\,${\upshape CC}}{\Indmm\Indmm
              \nextnr a$'$, SA $\gets$ {\sc Mutate}(e$'$, $\pi$, i, M$'$)\;
              \nextnr \lIf{{\upshape a$'$} {\upshape is not} {\upshape NULL}}{\Return{\upshape a$'$}}
              \nextnr a$'$ $\gets$ {\sc PickBestAssignment}($\pi$, SA)\;
              \nextnr \lIf{{\upshape a$'$} {\upshape is} {\upshape NULL}}{{\bf break}}
              \nextnr a $\gets$ a$'$\;
              \nextnr M$'$ $\gets$ {\sc FixInputBytes}(a, M$'$)\;
            }
         }
 \tikzmk{B}
 \boxit{gray2}
 \tikzmk{A}\nextnr \If{{\upshape opt}}{\Indmm
 \nextnr \lIf{{\upshape a} {\upshape is} {\upshape NULL}}{ a $\gets$ {\sc MutateOpt}(e, $\pi$, i, M)}
 \nextnr {\Return{\upshape a}}\;
 }
 \tikzmk{B}
 \boxit{gray2}
 \nextnr {\Return{\upshape NULL}}\
}
\vspace{2pt}
\caption{{\sc Solve} implementation of \fuzzysat: analysis stage in light gray, reasoning stage in dark gray (initial mutations due to $e$ at lines 3-4, multi-goal strategy at lines 5-15, and optimistic solving at lines 16-18).\label{alg:solve-one}}
\end{algorithm}
\setlength{\textfloatsep}{\textfloatsepsave}
\end{figure}

%% file: approach-analysis.tex
\subsection{Analyzing symbolic expressions}
\label{ss:analysis}

We now present the details of the main analyses integrated into the {\sc Analyze} function, which incrementally build the knowledge of \fuzzysat\ over an expression $e$.

\medskip\noindent
{\bf Detecting inputs and input groups.}
The first analysis identifies which input bytes $i_k$ are involved in an expression and evaluates how these bytes are grouped. In particular, \fuzzysat\ checks whether the expression can be regarded as an {\em input group}, i.e., the expression is equivalent to a concatenation ($\mdoubleplus$) of input bytes or constants that never {\em mix} their bits. Single byte expressions are also detected as input groups. 
\boxedspit{
\noindent{\bf\small Examples:}
\begin{itemize}
    \item expression $i_1 \mdoubleplus i_0$ contains inputs $i_0$ and $i_1$, and it is an input group since the bits from these bytes do not mix with each other but are just appended;
    \item expression $0 \mdoubleplus i_0$ contains input $i_0$ and it is a 1-byte input group as it is a zero-extend operation on $i_0$;
    \item expression $i_1 + i_0$ contains inputs $i_0$ and $i_1$, but it is not an input group as bits from $i_0$ are mixed, i.e., added, with bits from $i_1$;
    \item expression $(0 \mdoubleplus i_0) + (i_1 \ll 8)$ contains inputs $i_0$ and $i_1$, and it is an input group as the expression is equivalent to $i_1 \mdoubleplus i_0$, which is an input group.
\end{itemize}
}
Given an expression $e$, \fuzzysat\ stores in the expression metadata M the list of inputs involved in $e$, whether $e$ is an input group, and the list of input groups {\em contained} in $e$ when recursively considering subexpressions of $e$.

\medskip\noindent
{\bf Detecting uniquely defined inputs.} A crucial information about an input byte is knowing whether its value is fixed to a single value, dubbed {\em uniquely defined} in our terminology, due to one equality constraint that involves it. Indeed, it is not productive to fuzz input bytes whose value is fixed to a single constant. Given an expression $e$, then:
\begin{itemize}
    \item if $e$ is an equality constraint and one of its operands is an input group, or contains exactly only one input group, while the other operand is a constant, then M is updated to reflect that the bytes in the group will be uniquely defined due to $e$ if $e$ is later added to $\pi$;
    \item an input in $e$ is marked as uniquely defined whenever a constraint from $\pi$ marks it as uniquely defined;
    \item the input group in $e$ (if any) is marked as uniquely defined whenever the inputs forming it are all uniquely defined due to constraints in $\pi$;
\end{itemize}
\boxedspit{
\noindent{\bf\small Example.} The expression $i_1 \mdoubleplus i_0 == \mathrm{0xABCD}$ makes \fuzzysat\ mark inputs $i_0$ and $i_1$ as uniquely defined. If this expression is later added to $\pi$, then $i_0$ and $i_1$ will be considered uniquely defined in other expressions, disabling fuzzing on their values. 
}

\medskip\noindent
{\bf Detecting input-to-state branch conditions.} This analysis checks whether $e$ contains at least one operand that has input-to-state correspondence (\cref{se:related}). In \fuzzysat\, we use the following conditions to detect this kind of branch conditions: (a) $e$ matches the pattern $e'~op_{cmp}~e''$, where $op_{cmp}$ is a comparison operator (e.g., $\ge$, $==$, etc.) and (b) one operand ($e'$ or $e''$) is an input group. 
When $e$ is a Boolean negation, \fuzzysat\  recursively analyzes the subexpression. 
\boxedspit{
\noindent{\bf\small Example.} The expression $\mathrm{10} \ge i_1 \mdoubleplus i_0$ is an input-to-state branch condition as $\ge$ is a comparison operator and $i_1 \mdoubleplus i_0$ is an input group.
}

\medskip\noindent
{\bf Detecting interesting constants.} \fuzzysat\ checks the expression $e$, looking for constants that could be valuable during the reasoning stage, dynamically building a dictionary to use during the transformations. When specific patterns are detected, \fuzzysat\ generates variants of the constants based on the semantics of the computation performed by $e$.
\boxedspit{
\noindent{\bf\small Example.} When analyzing $i_1\,\oplus\,\mathrm{0xF0} == \mathrm{0x0F}$, \fuzzysat\ collects the constants $\mathrm{0xF0}$, $\mathrm{0x0F}$, and $\mathrm{0xFF}$ (i.e., $\mathrm{0xF0} \oplus \mathrm{0x0F}$) since the computation is an exclusive or. 
}
The patterns used to generate interesting constants can be seen as a relaxation of the concept of input-to-state relations.

\medskip\noindent
{\bf Detecting range constraints.} \fuzzysat\ checks whether $e$ is a {\em range constraint}, i.e., a constraint that sets a lower bound or an upper bound on the values that are admissible for the input group in $e$ (if any). For instance, \fuzzysat\ looks for constraints matching the pattern $e'~op_{cmp}~e''$ where $e'$ is an input group, $op_{cmp}$ is a comparison operator, and $e''$ is a constant value. Other equivalent patterns, such as $(e'~-~e'')~op_{cmp}~e'''$ where $e'$ is an input group while $e''$ and $e'''$ are constants, are detected as range constraints as well. 

By considering bounds resulting from expressions in $\pi$ and not only from $e$, \fuzzysat\ can compute refined range intervals for the input groups contained in an expression. To compactly and efficiently maintain these intervals, \fuzzysat\ uses {\em wrapped intervals}~\cite{wrapped-intervals} which can transparently deal with both signed and unsigned comparison operators.
\boxedspit{
\noindent{\bf\small Examples.} 
\begin{itemize}
    \item given the expressions $i_1 \mdoubleplus i_0 > \mathrm{10}$ and $i_1 \mdoubleplus i_0 \le 30$, \fuzzysat\ computes the range interval $[11, 30]$ for the input group composed by $i_0$ and $i_1$;
    \item given the expression $(i_1 \mdoubleplus i_0) + \mathrm{0xAAAA} <_{unsigned} \mathrm{0xBBBB}$, \fuzzysat\ computes the intervals $[0, \mathrm{0x1110}] \cup [\mathrm{0x5556}, \mathrm{0xFFFF}]$ for $i_0$ and $i_1$, correctly 
    modeling the wrap-around that may result in the two's complement representation.
\end{itemize}
}

\medskip\noindent
{\bf Detecting conflicting expressions.} The last analysis is devised to identify which expressions from $\pi$ may {\em conflict} with $e$ when assigning some of its input bytes. In particular, \fuzzysat\ marks an expression $e'$ as in conflict with $e$ whenever the set of input bytes in $e'$ is not disjoint with the set from $e$.
\boxedspit{
\noindent{\bf\small Example.} 
The expression $i_1 + i_0 > 10$ is in conflict with the expression $i_1 + i_2 < 20$ as they both contain the input byte $i_1$. Hence, fuzzing the first expression may negatively affect the second expression.

}
Computing the set of conflicting expressions is essential for performing the multi-goal strategy during the reasoning stage. 

%% file: approach-reasoning.tex
\subsection{Fuzzing symbolic expressions}
\label{ss:reasoning}

The core step during the reasoning stage of \fuzzysat\ is the execution of the function {\sc Mutate}, which attempts to find a valid assignment $a$. 
To reach this goal, {\sc Mutate} performs a sequence of mutations over the input test case $i$, returning as soon as a valid assignment is found by one of these transformations. When a mutation generates an assignment that satisfied $e$ but not $\pi$, then {\sc Mutate} saves it into a set of candidate assignment $SA$, which could be valuable later on during the multi-goal strategy (\cref{ss:overview}). In some cases, a transformation can determine that there exists a contradiction between $e$ and the conditions in $\pi$, leading {\sc Mutate} to an early termination. Additionally, when {\sc Mutate} builds a candidate assignment $a$, it checks that $a$ is consistent with the range intervals known for the modified bytes, discarding $a$ in case of failure and avoiding the (possibly expensive) check over $\pi$. We now review in detail the input transformations performed by the function {\sc Mutate}. 

\medskip\noindent
{\bf Fuzzing input-to-state relations.} When an expression $e$ is an input-to-state branch condition (\cref{ss:analysis}), \fuzzysat\ tries to replace the value from one operand $e'$ into the bytes composing the input group from the other (input-to-state) operand $e''$. If $e'$ is not constant, then \fuzzysat\ gets its concrete value by evaluating $e'$ on the test case $i$. When $e'$ is constant and the relation is an equality, if the assignment does not satisfy $\pi$, then \fuzzysat\ deems the query unsatisfiable. 
Conversely, when the comparison operator is not an equality, \fuzzysat\ tests variants of the value from $e'$, e.g., by adding or subtracting one to it, in the same spirit as done by \redqueen.
\vspace{-4mm}
\boxedspit{
\noindent{\bf\small Example.} Given $i_1 \mdoubleplus i_0 == \mathrm{0xABCD}$, \fuzzysat\ builds the assignment $\{i_0 \gets \mathrm{0xCD}, i_1 \gets \mathrm{0xAB}\}$. If the range interval over $i_0$ is $[\mathrm{0xDD}, \mathrm{0xFF}]$ due to constraints from $\pi$, then the assignment can be discarded without testing $\pi$, deeming the query unsatisfiable (but keeping the assignment in SA in case of optimistic solving).
}
\vspace{-1mm}

\medskip\noindent
{\bf Range interval brute force.} When a range interval is known for an input group contained an expression $e$, \fuzzysat\ can use this information to perform brute force on its value and possibly find a valid assignment. In particular, when an expression contains a single input group and its range interval is less than 2048, \fuzzysat\ builds assignments that brute force all the possible values assignable to the group. If no valid assignment is found, then the query can be deemed unsatisfiable. If the interval is larger than 2048, then \fuzzysat\ only tests the minimum and maximum value of the interval. To make this input transformation less conservative, \fuzzysat\ runs it even when $e$ contains at least one input group whose interval is less than\footnote{We pick the input group with the minimum range interval. 
} 512. 
\boxedspit{
\noindent{\bf\small Example.} Given the expression $(i_1 \mdoubleplus i_0) * \mathrm{0xABCD} == \mathrm{0xCAFE}$ and the range interval $[1, 9]$ (built due to constraints from $\pi$) on the group $g$ with $i_0$ and $i_1$, then \fuzzysat\ builds assignments for $g \in [1, 9]$, deeming the query unsatisfiable if none of them satisfies $e \wedge \pi$.
}

\medskip\noindent
{\bf Trying interesting constants.} For each constant $c$ collected by {\sc Analyze} when considering the expression $e$ and for each input group $g$ contained in $e$, \fuzzysat\ tries to set the bytes from $g$ to the value $c$. Since constants are collected through {\em relaxed} patterns, \fuzzysat\ tests different encodings (e.g., little-endian, big-endian, zero-extension, etc.) for each constant to maximize the chances of finding a valid assignment.
\boxedspit{
\noindent{\bf\small Example.} Given the expression $(i_1 \mdoubleplus i_0) * 100 == 200$ and assuming that {\sc Analyze} has collected the constants $\{2, 99, 100, 101, 199,200, 201\}$ where $2$ was obtained as $200/100$, while other constants are obtained from $100$ and $200$, then \fuzzysat\ would find a valid assignment when testing $\{i_0 \gets 2, i_1 \gets 0\}$ ($c = 2$, little-endian encoding).
\vspace{-0.5mm}
}

\medskip\noindent
{\bf Gradient descent.} Given an expression $e$, \fuzzysat\ tries to reduce the problem of finding a valid assignment for it to a minimization (or maximization) problem. This is valuable not only in the context of {\sc SolveMin}, {\sc SolveMax}, or {\sc SolveAll} where this idea seems natural, but also when reasoning over the branch condition $e$ in {\sc Solve}. Indeed, any expression of the form $e'\,op_{cmp}\,e''$, where $op_{cmp}$ is a comparison operator, can be transformed into an expression $f$ amenable to minimization to find a valid assignment~\cite{angora}, e.g.,  $e' < e''$ can be transformed\footnote{For the sake of simplicity, we ignore in our examples the wrap-around.} 
 into $f < 0$ with $f = e' - e''$. 

The search algorithm implemented in \fuzzysat\ is inspired by \angora~\cite{angora} and it is based on gradient descent. Although this iterative approach may fail to find a global minimum for $f$, a local minimum can be often {\em good enough} in the context of concolic execution as we do not always really need the global minimum but just an assignment that satisfies the condition, e.g., given $i_0 < 1$, the assignment $\{i_0 \gets \mathrm{0x0}\}$ satisfies the condition even if the global minimum for $i_0 - 1$ is given by $\{i_0 \gets \mathrm{0x81}\}$. For this reason, \fuzzysat\ in {\sc Solve} can stop the gradient descent as soon an assignment satisfies both $e$ and $\pi$. 
When the input groups from $e$ have disjoint bytes, \fuzzysat\ computes the gradient considering groups of bytes, instead of computing it for each distinct byte, as this makes the descent more effective. In fact, reasoning on $i_0$ and $i_1$ as a single value is more appropriate when these bytes are used in a two-byte operation since gradient descent may fail when these bytes are considered independently.

\boxedspit{
\noindent{\bf\small Example.} Given the expression $(i_0 \mdoubleplus i_1) - 10 > (i_2 \mdoubleplus i_3) - 5$ and a zero-filled input test case, then \fuzzysat\ transforms the expression into $((i_2 \mdoubleplus i_3) - 5) - ((i_1 \mdoubleplus i_0) - 10) < 0$, computes the gradients over the input groups $(i_1 \mdoubleplus i_0)$ and $(i_2 \mdoubleplus i_3)$, finding the assignment $\{i_0 \gets \mathrm{0x80}, i_1 \gets \mathrm{0x06}, i_2 \gets \mathrm{0x84}, i_3 \gets \mathrm{0x01}\}$ which 
makes the condition satisfied as $(\mathrm{0x80} \mdoubleplus \mathrm{0x06}) - 10 = 32764 > -31748 = (\mathrm{0x84} \mdoubleplus \mathrm{0x01}) - 5$.
\vspace{-0.5mm}
}
\vspace{-1mm}

\medskip\noindent
{\bf Deterministic and non-deterministic mutations.} These two sets of input transformations are inspired by the two mutation stages from \afl\ (\cref{se:related}). Deterministic mutations include bit or byte flips, replacing bytes with interesting {\em well-known} constants (e.g., {\tt MAX\_INT}), adding or subtracting small constants from some input bytes. Non-deterministic mutations instead involve also transformations such as flipping of random bits inside randomly chosen input bytes, setting randomly chosen input bytes to randomly chosen interesting constants, subtracting or adding random values to randomly chosen bytes, and several others~\cite{AFL}.
%
%
%
%
The main differences with respect to \afl\ are: (a) mutations are applied only on the input bytes involved in the expression $e$, (b) multi-byte mutations are considered only in the presence of multi-byte input groups, (c) for non-deterministic mutations, \fuzzysat\ generates $k$ distinct assignments, with $k$ equal to $\max\{100,n_{i}\cdot 20\}$ where $n_i$ is the number of inputs involved in $e$, and for each assignment it applies a sequence (or stack) of $n$ mutations ($n = 1 \ll (1 + rand(0,7))$ as in \afl).

%% file: implem.tex
\section{Implementation}
\label{se:implem}

\fuzzysat\ is written in C (10K LoC) and evaluates queries in the language used by the Z3 Theorem Prover~\cite{Z3-TACS08}. To efficiently evaluate an expression given a concrete assignment 
\fuzzysat\ uses a fork of Z3 where  the {\tt Z3\_model\_eval} function has been optimized to deal with full concrete models.

\fuzzolic\ is a new concolic executor based on {\sc QEMU} 4.0 (user-mode), written in C (20K LoC), that currently supports Linux x86\_64 binaries. Its design overcomes one of the major problems affecting \qsym: \fuzzolic\ decouples the tracer component, which builds the symbolic expressions, from the solving component, which reasons over them. This is required as recent releases of most DBI frameworks, such as {\sc PIN}~\cite{Pin-PLDI05} on which \qsym\ is based on, do not allow an analysis tool to use external libraries (as the Z3 solver in case of \qsym) when they may produce side effects on the program under analysis~\cite{DCNPC-ASIACCS19}. This implementation constraint has made it very complex to port QSYM to newer releases of PIN, limiting its compatibility with recent software and hardware configurations\footnote{QSYM has been recently removed from the Google project FuzzBench due to its instability on recent Linux releases~\cite{qsym-removed-fuzzbench}.}. To overcome this issue, the two components are executed into distinct processes in \fuzzolic. In particular, the tracer runs under QEMU and generates symbolic expressions in a compact language, storing them into a shared memory that is also attached to the memory space of the solving component, which in turn 
submits queries to \fuzzysat\ to generate alternative inputs. Similarly to \qsym, \fuzzolic\ runs in parallel with two coverage-guided fuzzers.


%% file: eval.tex
\section{Evaluation}
\label{se:evaluation}

\noindent In this section we 
address the following research questions:

\begin{itemize}

\item {\bf RQ1}: How effective and efficient is \fuzzysat\ at solving queries generated by concolic executors?
\item {\bf RQ2}: How do different kinds of mutations help \fuzzysat\ in solving queries?
\item {\bf RQ3}: How does \fuzzolic\ with \fuzzysat\ compare to state-of-the-art fuzzers on real-world programs?

\end{itemize}

{\bf Benchmarks.} Throughout our evaluation, we consider the following 12 programs: advmng 2.00, bloaty rev {\tt 7c6fc}, bsdtar rev. {\tt f3b1f}, djpeg v9d, jhead 3.00-5, libpng 1.6.37, lodepng-decode rev. {\tt 5a0dba}, objdump 2.34, optipng 0.7.6, readelf 2.34, tcpdump 4.9.3 (libpcap 1.9.1), and tiff2pdf 4.1.0. These targets have been heavily fuzzed by the community~\cite{google-oss-fuzz}, and used in previous evaluations of state-of-the-art fuzzers~\cite{SYMCC,QSYM,ECLIPSER-ICSE19,WEIZZ,redqueen}. As seeds, we use the AFL test cases~\cite{AFL}, or when missing, 
minimal syntactically valid files~\cite{small-files-project}.


\vspace{0.2mm}
{\bf Experimental setup.} We ran our tests in a Docker container based on the Ubuntu 18.04 image, using a server with two Intel Xeon E5-4610v2@2.30 GHz CPUs and 256 GB of RAM. 

\subsection{RQ1: Solving effectiveness of \fuzzysat}

To evaluate how effective and efficient is \fuzzysat\ at solving queries generated by concolic execution, we discuss an experimental comparison of \fuzzysat\ against the SMT solver Z3 and the approximate solver \jfs. We first focus on {\sc Solve} queries, collected by running the 12 programs under QSYM on their initial seed with optimistic solving disabled, comparing the solving time and the number of queries successfully proved as satisfiable when using these three solvers. Then, we analyze the performance of \qsym\ at finding bugs on the LAVA-M dataset~\cite{LAVAM} when using \fuzzysat\ with respect to when using Z3, implicitly considering the impact also of other reasoning primitives (e.g., {\sc SolveMax}) and from enabling optimistic solving in {\sc Solve}. In these experiments, we consider \qsym\ instead of \fuzzolic\ to avoid any bias resulting from its expression generation phase that could benefit \fuzzysat\ and impair the other solvers.


\input{table-fuzzy-vs-z3}

\medskip
{\bf \bfseries{{\scshape Fuzzy-Sat} vs Z3}.}
Table~\ref{tab:z3-vs-fuzzy} provides an overview of the comparison between \fuzzysat\ and Z3 on the queries generated 
when running the 12 benchmarks. 

The first interesting insight is that only a small subset of the queries, i.e., less than 10\%, has been proved satisfiable (even when considering together both solvers). The remaining queries are either proved unsatisfiable or make the solvers run out of the time budget (10 seconds for Z3, as in \qsym). 

The second insight is that, when focusing on the queries that are satisfiable, \fuzzysat\ is able to solve the majority of them and can even perform better than Z3 on a few benchmarks: for instance, \fuzzysat\ solves $301$ ($236.7+64.3$) queries on average on advmng, while Z3 stops at $243.7$ ($236.7+7$). Although this may seem unexpected, this result is consistent with past evaluations from state-of-the-art fuzzers~\cite{ECLIPSER-ICSE19,redqueen} that have shown that a large number of branch conditions 
can be solved even without SMT solvers. Nonetheless, there are still a few queries were \fuzzysat\ is unable to find a valid assignment while Z3 is successful, e.g., \fuzzysat\ misses 7 queries on bloaty (but solves one query that makes Z3 run out of time). Assessing the impact of {\em solving} or {\em not solving} a query in concolic execution is a hard problem, especially when bringing into the picture hybrid fuzzing and its non-deterministic behavior. Hence, we only try to indirectly speculate on this impact by later discussing the results on the LAVA-M dataset and the experiments in Section~\ref{ss:eval-coverage}.


Lastly, we can see in Table~\ref{tab:z3-vs-fuzzy} that on average \fuzzysat\ requires $31\times$ less time than Z3 to reason over the queries from the 12 benchmarks. When putting together this result with the previous experimental insights, we could speculate why \fuzzysat\ could be beneficial in the context of concolic execution: it can significantly reduce the solving time during the concolic exploration while still be able to generate a large number of (possibly valuable) inputs.

\begin{figure}[!t]
    \centering
    \begin{minipage}{.03\textwidth}
        \centering
        {\scriptsize (a)}\hspace*{-5mm}
    \end{minipage}%
    \begin{minipage}{.23\textwidth}
        \centering
        \includegraphics[width=0.75\linewidth]{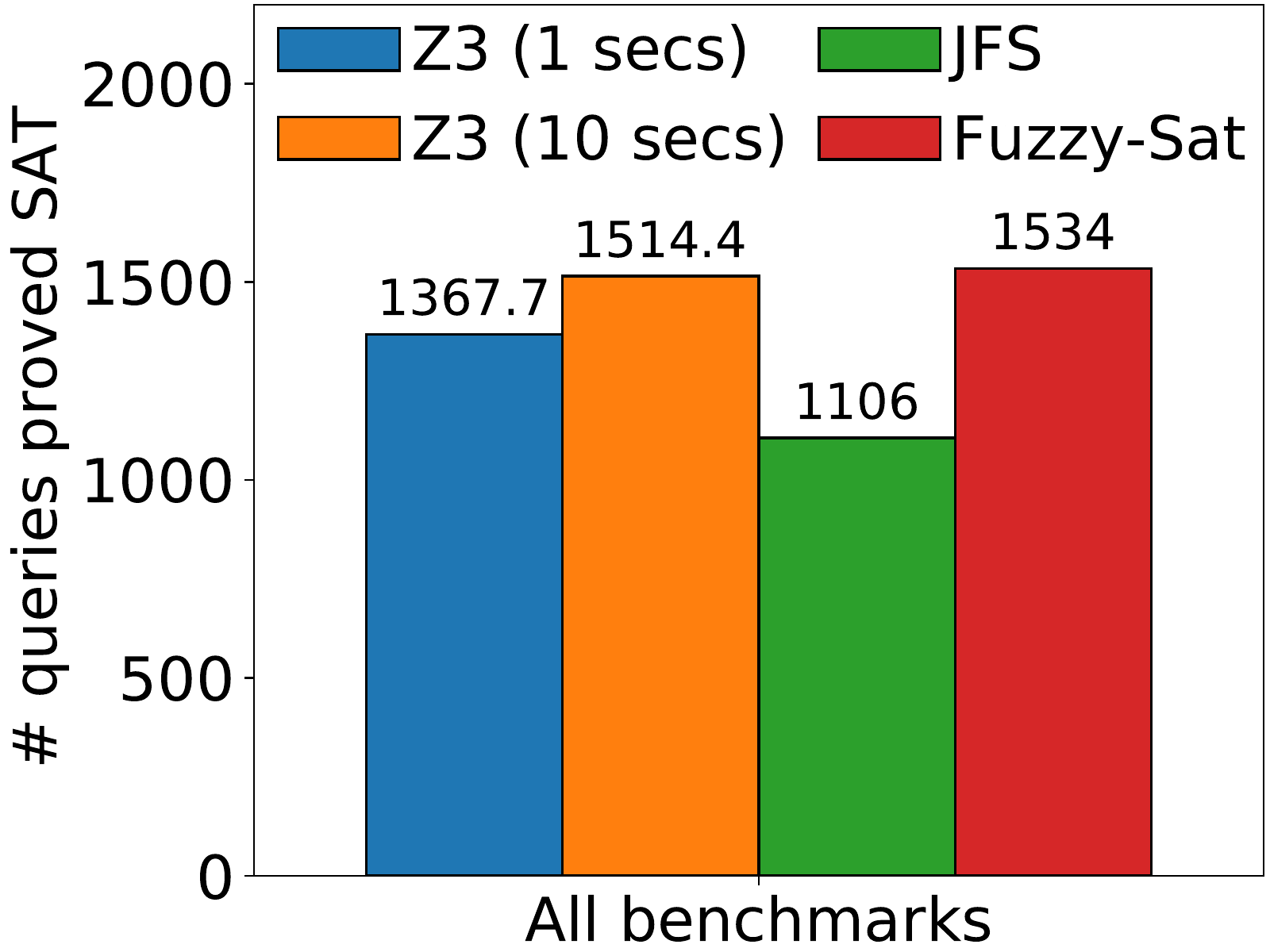}
    \end{minipage}%
    \begin{minipage}{.03\textwidth}
        \centering
        {\scriptsize (b)}\hspace*{-5.0mm}
    \end{minipage}%
    \begin{minipage}{0.23\textwidth}
        \centering
        \includegraphics[width=0.75\linewidth]{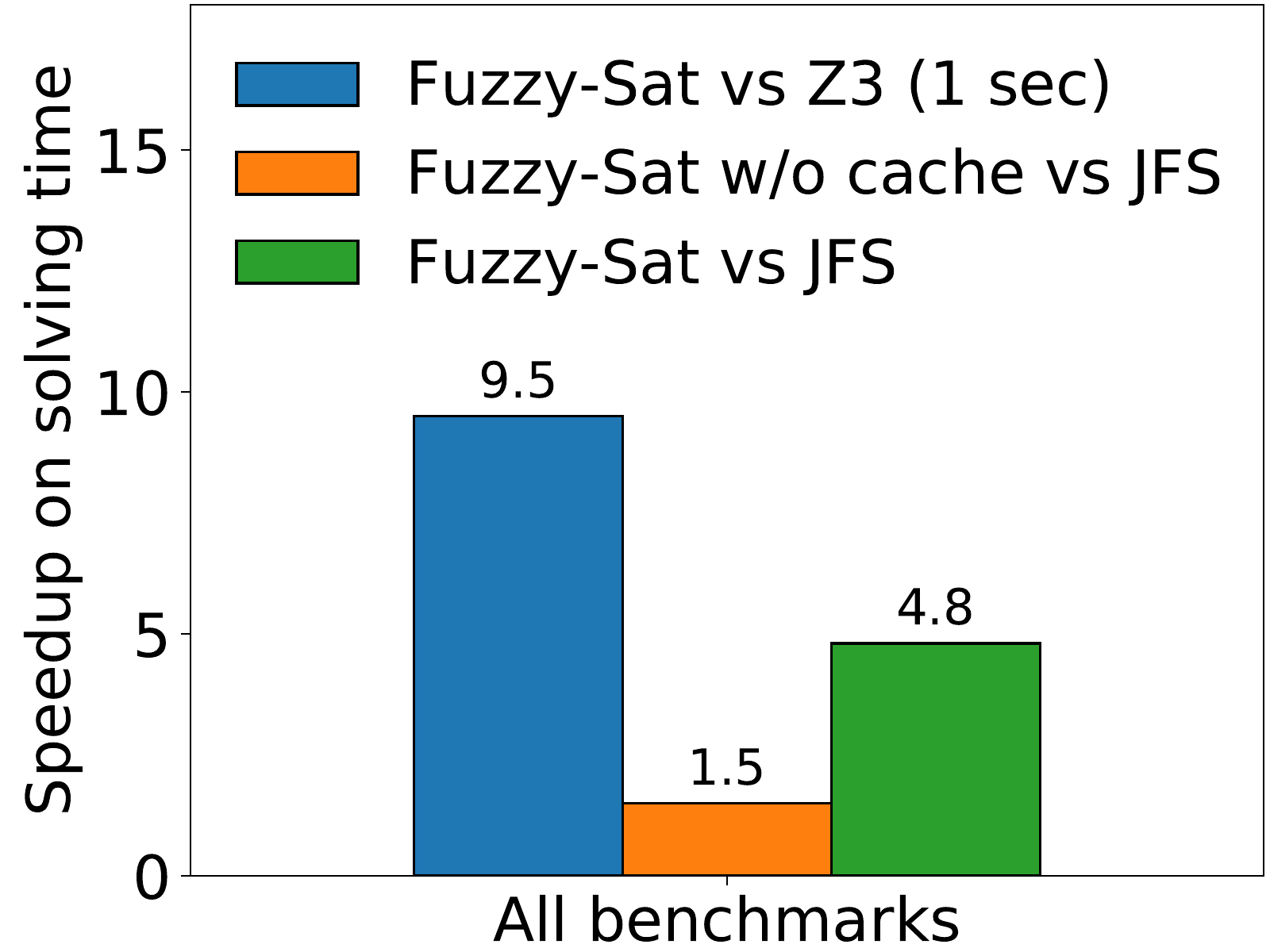}
    \end{minipage}
    \vspace{-3mm}
    \caption{\fuzzysat\ vs other solvers on the 12 benchmarks: (a) number of queries proved satisfiable and (b) speedup on the solving time.\label{fig:fuzzy-sat-vs-z3-jsf}}
\end{figure}

One natural question is whether one could get the same benefits of \fuzzysat\ by drastically reducing the time budget given to Z3. To tackle this observation, Figure~\ref{fig:fuzzy-sat-vs-z3-jsf}a reports the number of queries solved by Z3 when using a timeout of 1 second and Figure~\ref{fig:fuzzy-sat-vs-z3-jsf}b shows how the speedup from \fuzzysat\ is reduced in this setup. \fuzzysat\ is still $9.5\times$ faster than Z3 and the gap between the two 
in terms of solved queries increases significantly ($+12\%$ in \fuzzysat), suggesting that this setup of Z3 is not as effective as one may expect.

\medskip
{\bf \bfseries{{\scshape Fuzzy-Sat} vs {\scshape JSF}.}} One solver that shares the same spirit of \fuzzysat\ is \jfs\ (\cref{se:related}), which however is based on a different design. When considering the queries collected on the 12 benchmarks, it can be seen in Figure~\ref{fig:fuzzy-sat-vs-z3-jsf}a that \jfs\ is able to solve only 1106 queries, significantly less than the 1534 from \fuzzysat. On 127 out of the 325 queries from bsdtar, \jfs\ has failed to generate the program to fuzz due to the large number of nested expressions contained in the queries, 
yielding a gap of $95$ solved queries between two solvers. The remaining missed queries can be likely explained by considering that: (a) it is not currently possible to provide the input test case $i$ used for generating the queries to the fuzzer executed by \jfs~\cite{jfs-custom-seeds}, as \jfs\ generates a program that takes an input that is different (in terms of size and structure) from $i$ and builds its own set of seeds, (b) \jfs\ does not provide specific insights to the fuzzer on how to mutate the input, and (c) \jfs\ uses {\sc LibFuzzer}~\cite{libfuzzer}, which does not integrate several fuzzing techniques that have inspired \fuzzysat.  

When considering the solving time, \fuzzysat\ is $1.5\times$ faster than \jfs\ (Figure~\ref{fig:fuzzy-sat-vs-z3-jsf}b). However, 
when enabling analysis cache in \fuzzysat, the speedup increases up to $4.8\times$.

\jfs\ does not currently provide a C interface~\cite{jfs-c-interface}, requiring concolic executors to dump the queries on disk: as this operation can take a long time in presence of large queries, we do not consider \jfs\ further in the other experiments.

\smallskip
{\bf \bfseries{{\scshape Fuzzy-Sat} on LAVA-M.}} %
%
To test 
whether \fuzzysat\ can solve queries that are valuable for a concolic executor,
we repeated the experiment on the LAVA-M dataset from the \qsym\ paper~\cite{QSYM}, 
looking for bugs within the four benchmarks base64, md5sum, uniq, and who. Table~\ref{tab:fuzzy-lava-m} reports the average and max number of bugs found during 5-hour experiments across 5 runs. \qsym\ with \fuzzysat\ finds on average more bugs than \qsym\ with Z3 on 3 out of 4 programs. In particular, the improvement is rather significant on who, where \fuzzysat\ allows \qsym\ to find $3\times$ more bugs compared to Z3, suggesting that trading performance for accuracy can be valuable in the context of hybrid fuzzing.

Interestingly, \fuzzysat\ was able to reveal bugs that the original authors from LAVA-M were unable to detect~\cite{LAVAM}, 
e.g., \fuzzysat\ has revealed 136 new bugs on who. Since other works~\cite{ECLIPSER-ICSE19,redqueen} reported a similar experimental observation, the additional bugs are likely not false positives.


\input{table-lava-m}

\subsection{RQ2: Impact of different kinds of mutations in \fuzzysat}
\label{ss:mutation-stats}

\input{table-fuzzy-stats}

An interesting question is which mutations contribute at making \fuzzysat\ effective. Table~\ref{tab:fuzzy-mutations-stats} reports which transformations have been crucial to solve the queries from the 12 benchmarks
, assigning a query to the multi-goal strategy when \fuzzysat\ had to reason over conflicting expressions from $\pi$ to solve the query. \fuzzysat\ was able to solve more than $51\%$ of the queries by applying input-to-state transformations, and an additional $17\%$ was solved by exploiting the interesting constants collected during the analysis stage. Range interval brute-force was helpful on around $10\%$ of the queries, while mutations inspired by AFL were beneficial in $8\%$ of them. Gradient descent solved just $1.5\%$ of the queries. However, two considerations must be taken into account: (a) the order of the mutations affect these numbers, as gradient descent is not used when previous (cheaper) mutations are successful, and (b) gradient descent is crucial for solving queries in {\sc SolveMin}, {\sc SolveMax}, and {\sc SolveAll}, which are not considered in this experiment. 
Finally, the multi-goal strategy of \fuzzysat\ was essential for solving around $11\%$ of the queries.

\subsection{RQ3: \fuzzysat\ in \fuzzolic}
\label{ss:eval-coverage}

\begin{figure*}[!t]
    \centering
    \begin{minipage}{.245\textwidth}
        \centering
        \includegraphics[width=0.99\linewidth]{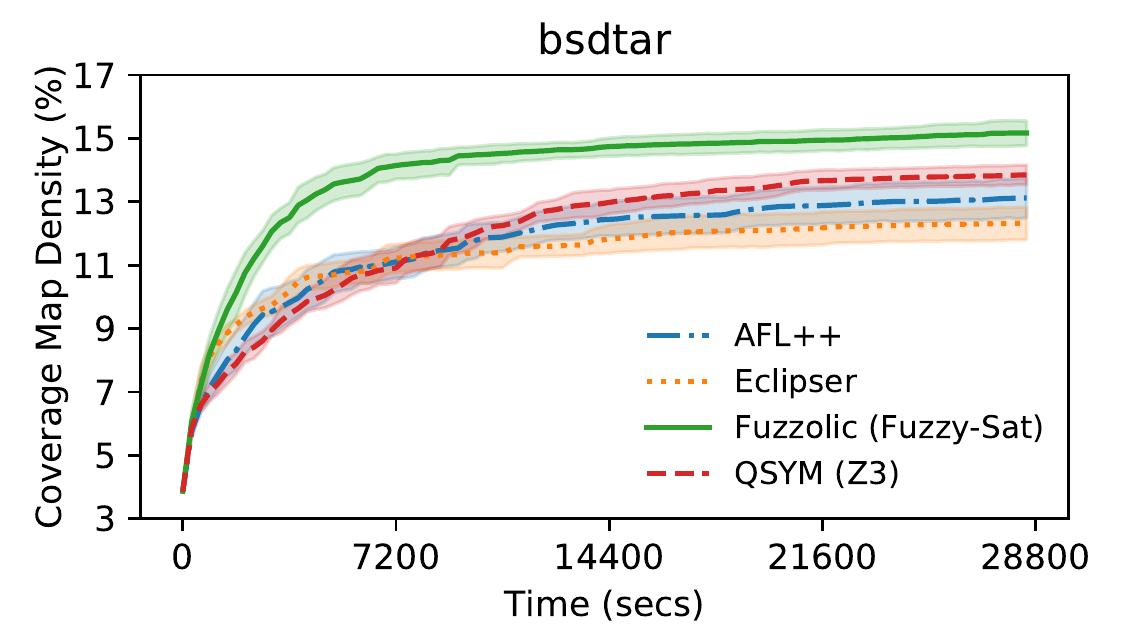}
    \end{minipage}%
    \begin{minipage}{0.245\textwidth}
        \centering
        \includegraphics[width=0.99\linewidth]{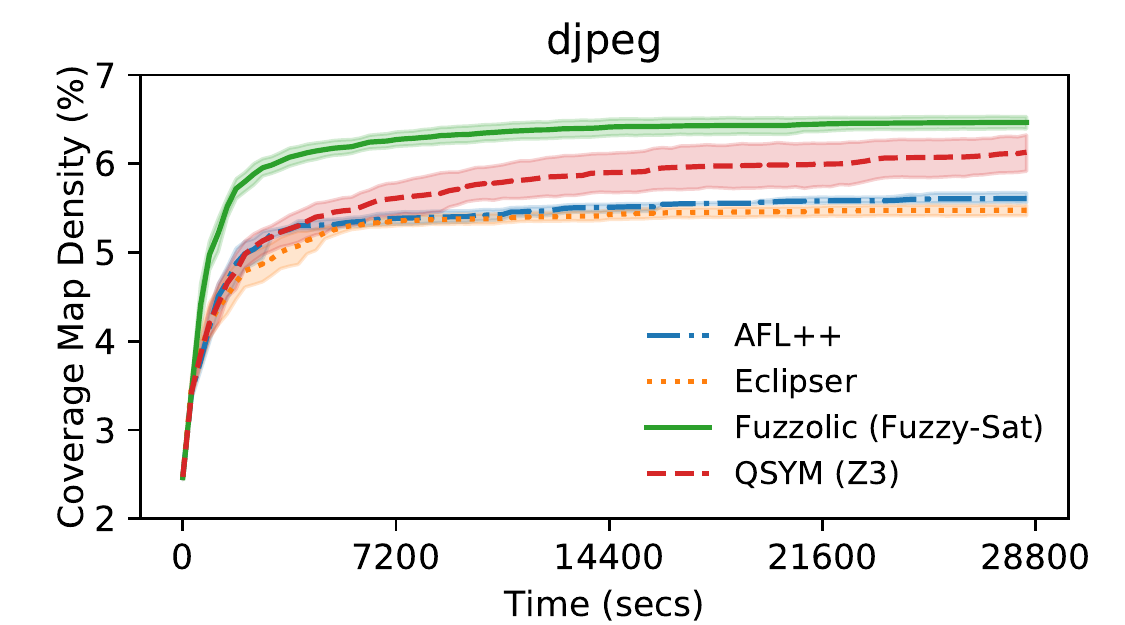}
    \end{minipage}
    \begin{minipage}{0.245\textwidth}
        \centering
        \includegraphics[width=0.99\linewidth]{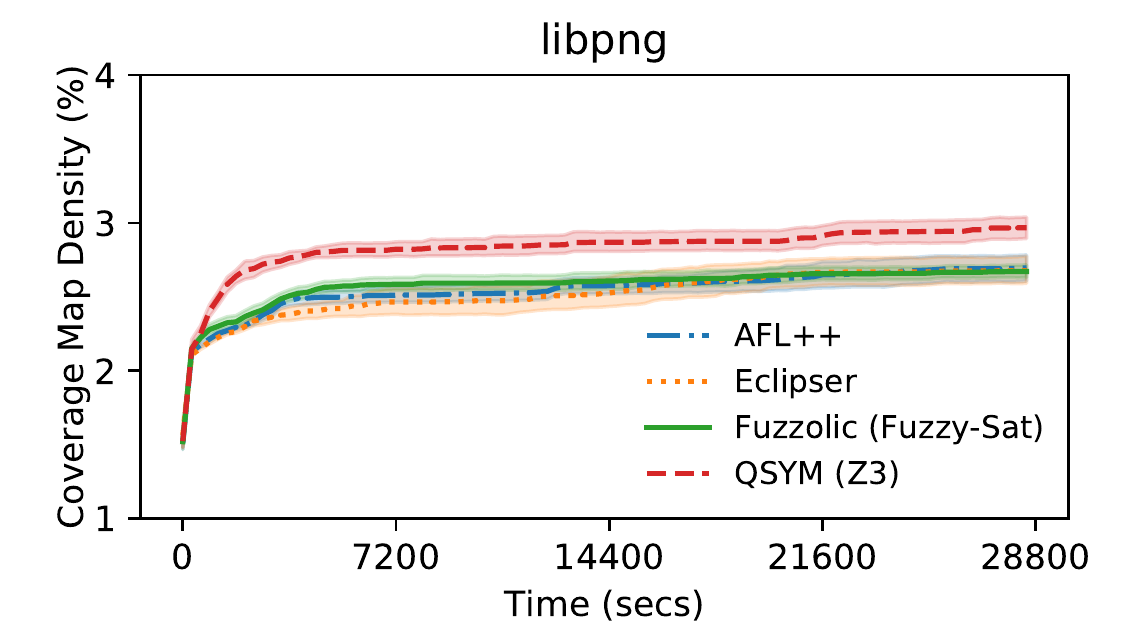}
    \end{minipage}
    \begin{minipage}{0.245\textwidth}
        \centering
        \includegraphics[width=0.99\linewidth]{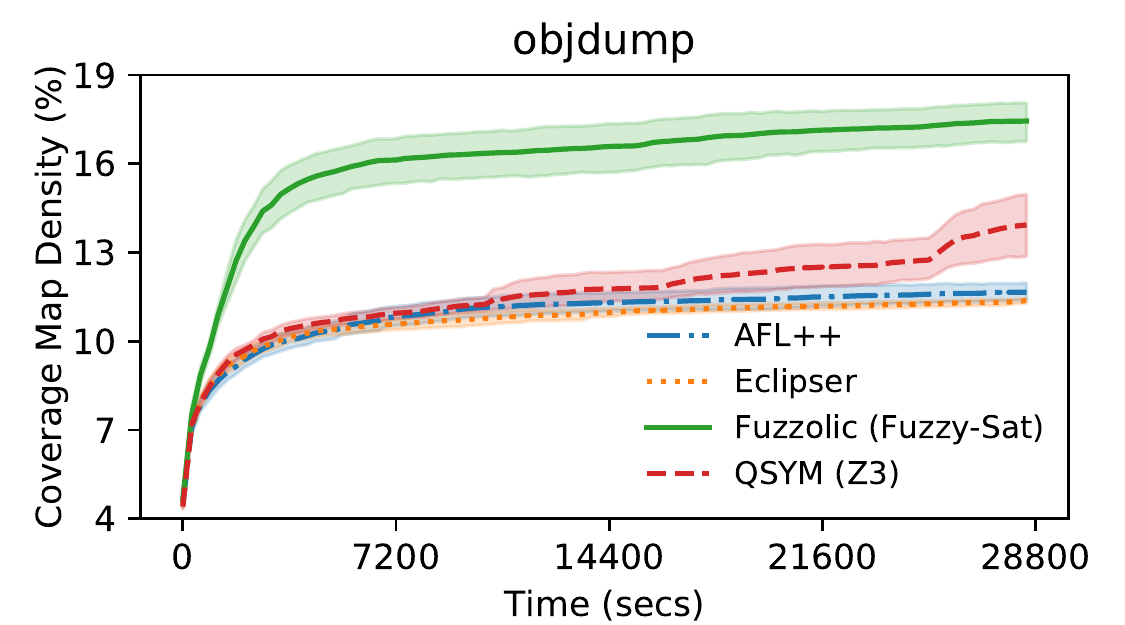}
    \end{minipage}
    \\\vspace{-2mm}
    \begin{minipage}{.245\textwidth}
        \centering
        \includegraphics[width=0.99\linewidth]{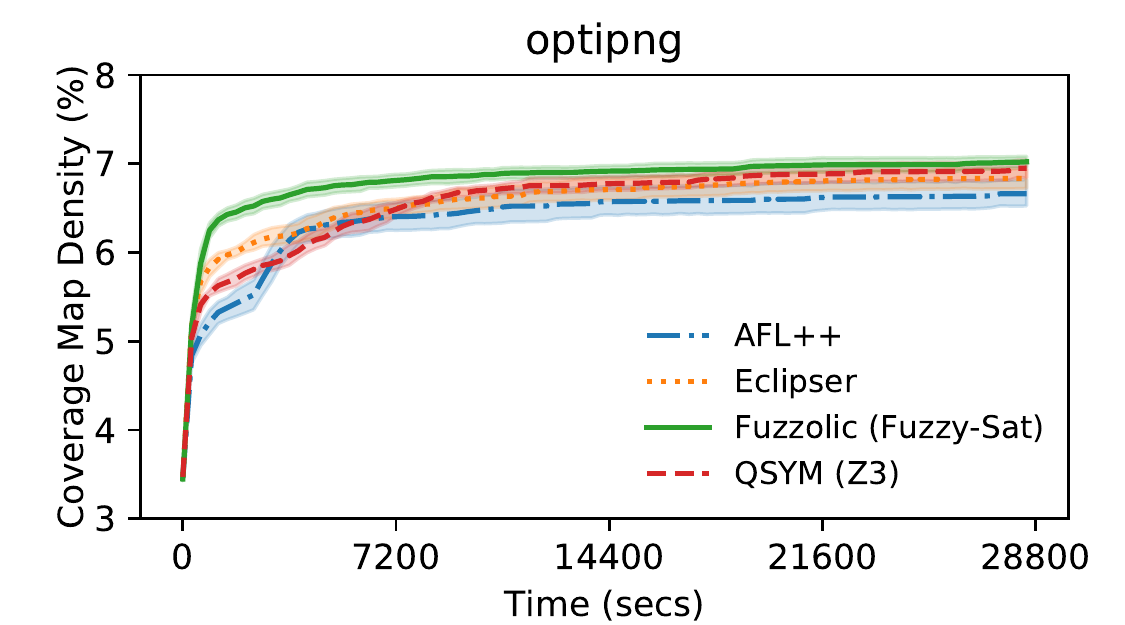}
    \end{minipage}%
    \begin{minipage}{0.245\textwidth}
        \centering
        \includegraphics[width=0.99\linewidth]{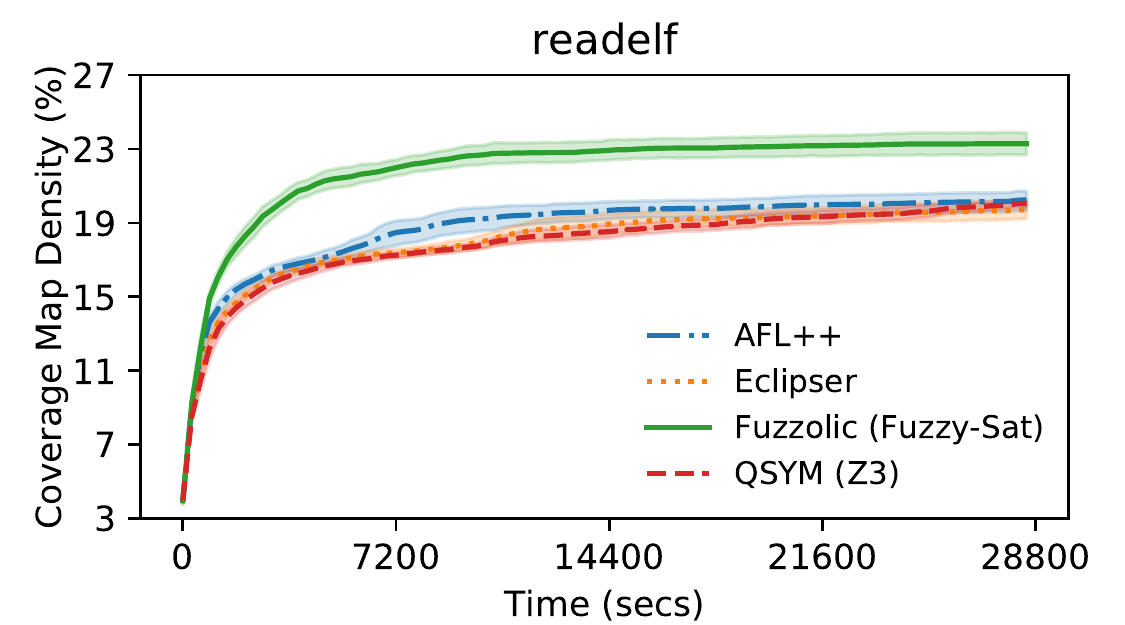}
    \end{minipage}
    \begin{minipage}{0.245\textwidth}
        \centering
        \includegraphics[width=0.99\linewidth]{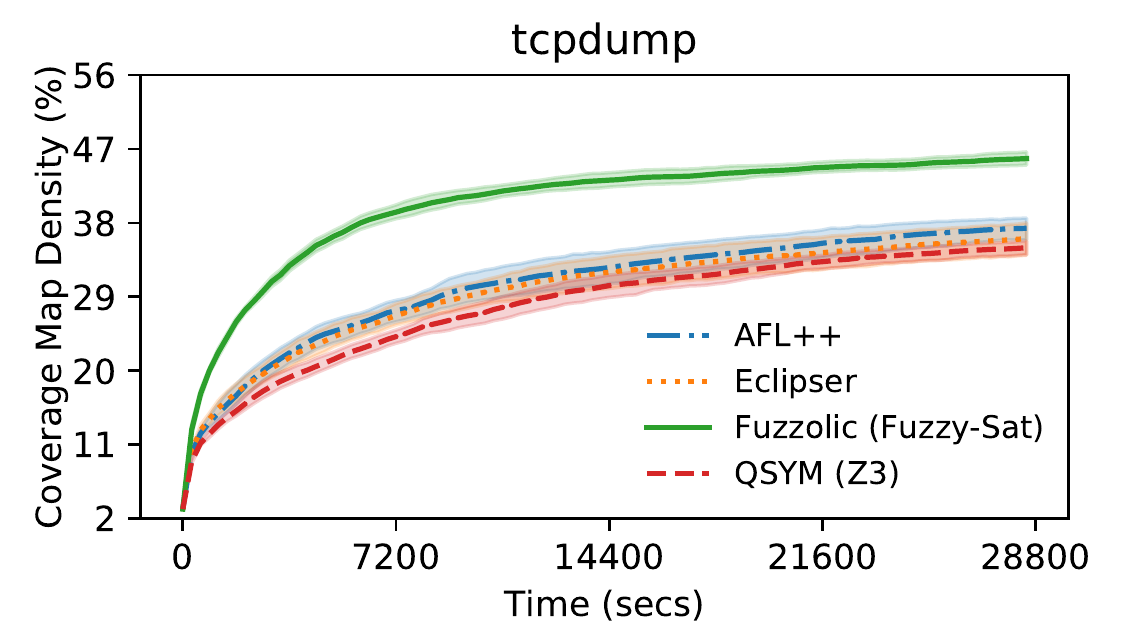}
    \end{minipage}
    \begin{minipage}{0.245\textwidth}
        \centering
        \includegraphics[width=0.99\linewidth]{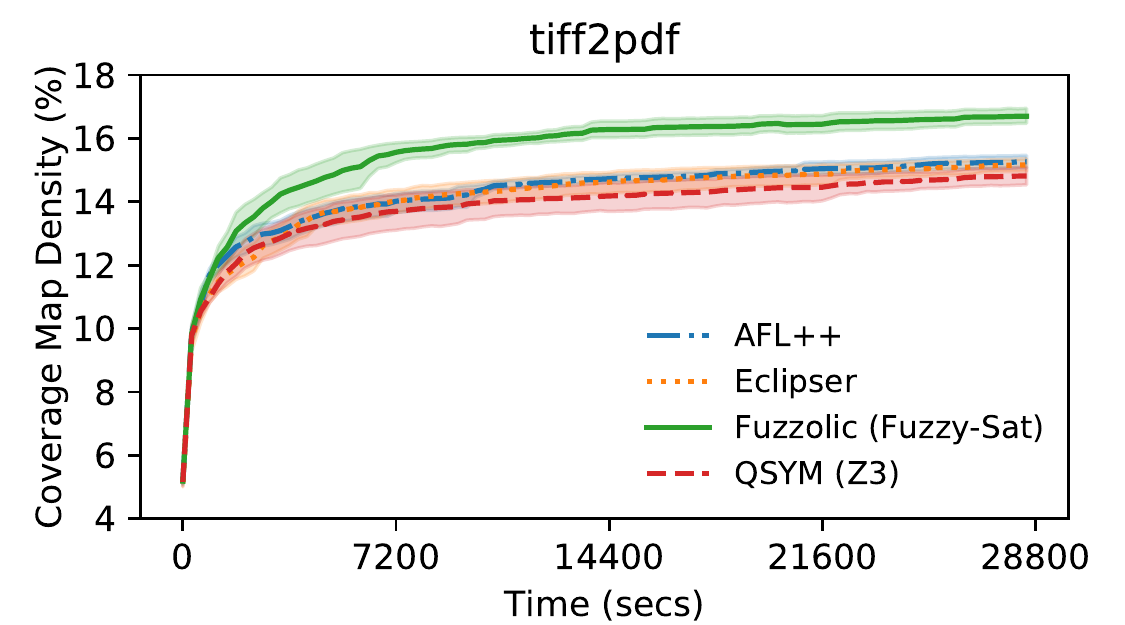}
    \end{minipage}
    \caption{Coverage map density 
    reached by \fuzzolic\ with \fuzzysat\ vs other state-of-the-art fuzzers. The shaded areas are the 95\% confidence intervals.\label{fig:coverage-fuzzolic}} 
\end{figure*}

To further assess the effectiveness of \fuzzysat, we compare \fuzzolic, which is built around this solver, against state-of-the-art binary open-source fuzzers on the 12 benchmarks, tracking the code coverage reached during 8-hour experiments (10 runs). Besides \fuzzolic, we consider: (a) \aflpp~\cite{AFLPP} rev. {\tt 3f128} in QEMU mode, which integrates~\cite{aflpp-colorization} the colorization technique from \redqueen, as well as other improvements to \afl\ proposed by the fuzzing community during the last years~\cite{AFLPP-WOOT20}, (b) \eclipser\ rev. {\tt b072f}, 
which devises one of the most effective {\em approximations} of concolic execution in literature, and (c) \qsym\ rev. {\tt 89a76} with Z3. As both \fuzzolic\ and \qsym\ are hybrid fuzzers that are designed to run in parallel with two instances ($F_m$, $F_s$) of a coverage-guided fuzzer, we consider for a fair comparison \aflpp\ and \eclipser\ in a similar setup, running them in parallel to ($F_m$, $F_s$) and allowing the tools to sync their input queues~\cite{aflpp-parallel-mode}. Hence, each run takes $8\times3=24$ CPU hours. For $F_m$ we use \aflpp\ in {\em master mode}, which performs deterministic mutations, while for $F_s$ we use \aflpp\ in {\em slave mode} that only executes non-deterministic mutations. Since \eclipser\ does not support a parallel mode, we extended it to allow \aflpp\ to correctly pick inputs from its queue. 


Figure~\ref{fig:coverage-fuzzolic} shows the code coverage reached by the different fuzzers on 8 out of 12 programs. On the remaining four benchmarks, the fuzzers reached soon a very similar coverage, making it hard to detect any significant trend and thus we omit their charts due to lack of space. Similar to other works~\cite{QSYM,SYMCC}, we plot the density of the coverage map from $F_s$ and depict the 95\% confidence interval using a shaded area. As bitmap collisions may occur~\cite{COLL-AFL}, we validated the trends by also computing the number of basic blocks~\cite{AFL-COV}. 


\begin{figure}[!t]
    \centering
    \begin{minipage}{.245\textwidth}
        \centering
        \includegraphics[width=0.99\linewidth]{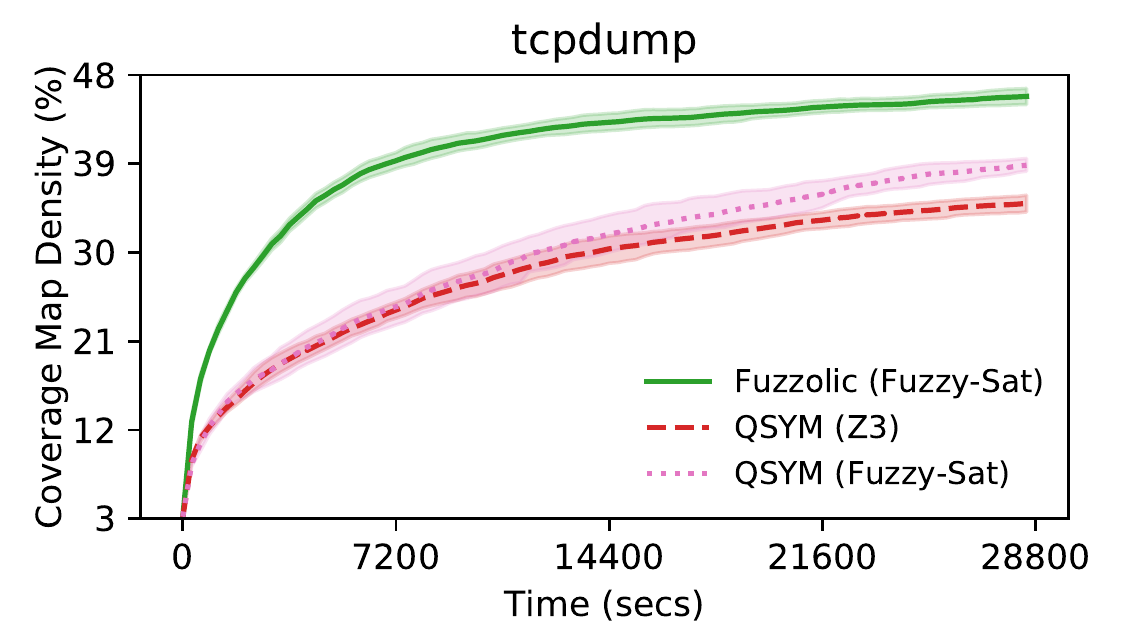}
    \end{minipage}%
    \begin{minipage}{0.245\textwidth}
        \centering
        \includegraphics[width=0.99\linewidth]{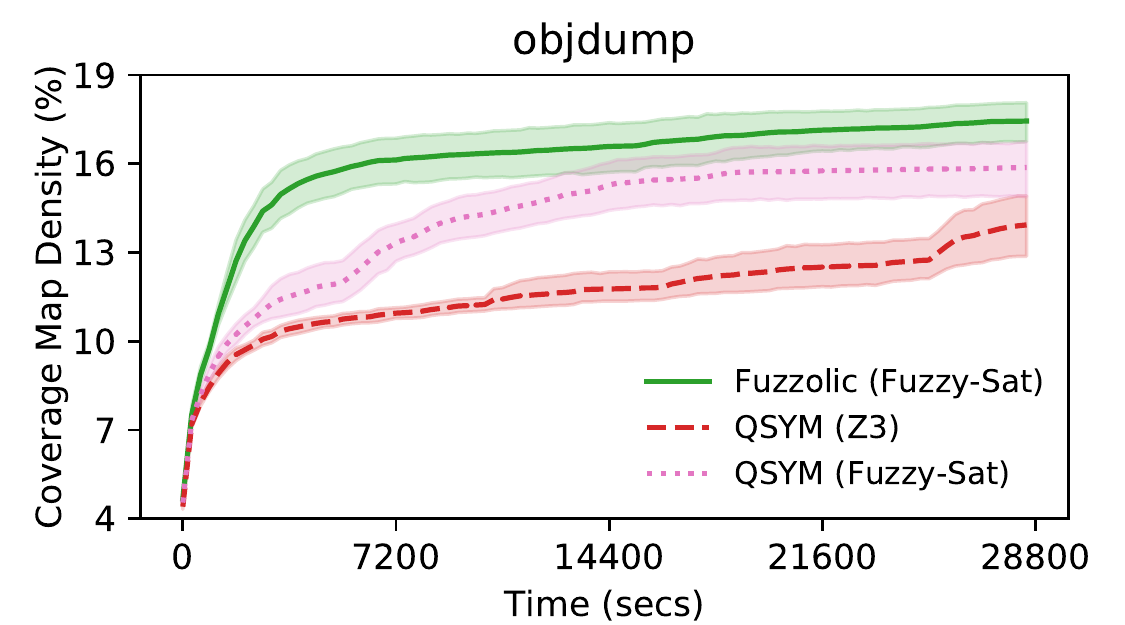}
    \end{minipage}
    \caption{Coverage map density
    : impact of \fuzzysat\ in \qsym.\label{fig:coverage-qsym-symcc}}
\end{figure}

\fuzzolic\ reaches a higher code coverage than other solutions on 6 programs, i.e., bsdtar, djpeg, objdump, readelf, tcpdump, and tiff2pdf. In particular, tcpdump is the program where \fuzzolic\ shines better, consistently showing over time an increase in the map density of $5\%$ with respect to the second-best fuzzer (\aflpp).  On optipng, although \fuzzolic\ appears to have an edge at the beginning of the experiment, it then reaches a coverage that is comparable to other fuzzers, which are all very close in performance. Finally, \fuzzolic\ falls behind other approaches on libpng, pinpointing one case where \fuzzysat\ seems to underperform compared to Z3, as \qsym\ dominates on this benchmark. 


When comparing closely \fuzzolic\ to \qsym, the improvement in the coverage is likely due to the better scalability of the former with respect to the latter. For instance, on tcpdump \fuzzolic\ performs concolic execution on $11089$ inputs ($1.8$ secs/input), generating $14415$ alternative inputs, while \qsym\ only analyzes $376$ inputs ($71.4$ secs/input) and generates $786$ alternative inputs. When considering libpng, \fuzzolic\ is still faster than \qsym\ ($8.8$ secs/input vs $43.6$ secs/input) but the number of inputs available in the queue from $F_s$ (from which \fuzzolic\ and \qsym\ pick inputs) over time is very low. 
Hence, the difference between \fuzzolic\ and \qsym\ on libpng is due to a few but essential queries that Z3 is able to solve while \fuzzysat\ fails to reason on.

The better scalability of \fuzzolic\ with respect to \qsym\ is given by the combination of an efficient solver (\fuzzysat) and an efficient tracer (\fuzzolic). Indeed, when replacing Z3 with \fuzzysat\ in \qsym, this concolic executor improves its performance but still falls behind \fuzzolic. Figure~\ref{fig:coverage-qsym-symcc} compares the coverage of \qsym\ with the two solvers and \fuzzolic\ on two benchmarks. On tcpdump, \qsym\ with \fuzzysat\ is able to analyze $2111$ inputs ($10.1$ secs/input) and generate $3690$ alternative inputs, improving the coverage by 4$\%$ on average with respect to Z3 but still performing worse than \fuzzolic. Similarly, on objdump the improvement in \qsym\ due to \fuzzysat\ is even more noticeable
 but still \qsym\ cannot match the coverage reached by \fuzzolic.


When comparing \fuzzolic\ to \eclipser\ and \aflpp, the results suggest that 
the integration of fuzzing techniques into a solver provides a positive impact. Indeed, while these fuzzers scales better than \fuzzolic, processing hundreds of inputs per second, they lack the knowledge that \fuzzysat\ extracts from the symbolic expressions, which is used to perform effective mutations. Overall, colorization from \aflpp\ and approximate concolic execution from \eclipser\ seem to generate similar inputs on several benchmarks, yielding often a similar coverage in our parallel fuzzing setup.
Moreover, despite \fuzzolic\ may spend several seconds over a single input, it collects information that allows it to fuzz a large number of branch conditions, paying on average only a few microseconds when testing an input assignment. Hence, the time spent building the symbolic expressions can be amortized over thousands of (cheap) query evaluations, reducing the gap between the efficiency of a fuzzer and a concolic executor. Nonetheless, \fuzzolic\ is still a hybrid fuzzer and it needs to run in parallel to a traditional fuzzer to provide good results, since some non-deterministic mutations, such as randomly combining inputs, are not performed by \fuzzolic.

%% file: table-fuzzy-vs-z3.tex

\begin{table}[t]
  \centering
  \caption{Number of queries proved satisfiable by \fuzzysat\ w.r.t. {\sc Z3} (timeout 10 secs). Numbers show the average of 5 runs. The speedup considers the solving time on the full set of  queries. \label{tab:z3-vs-fuzzy}} %
  \vspace{-1.5mm}
  \begin{adjustbox}{width=0.97\columnwidth,center}
  \begin{tabular}{ |l||c|c|c|c|c||c| }
    \hline
    \multirow{2}{*}{{\sc Program}} 
    & {\sc \#}
    & \multicolumn{3}{c|}{{\sc \# queries proved SAT by}} 
    & {\sc \# SAT \fuzzysat}  
    & {\sc Solv. Time}
    \\\cline{3-5}

    & {\sc queries} 
    & {\sc both} & {\sc Z3} & {\sc } \fuzzysat 
    & {\sc Div. by \# SAT Z3} 
    & {\sc Speedup}\\
    \hline\hline

advmng & 1481 & 236.7 & +7.0 & +64.3 & 1.24 & 17.1$\times$ \\ 
bloaty & 2085 & 95.0 & +7.0 & +1.0 & 0.94 & 47.8$\times$ \\ 
bsdtar & 325 & 124.0 & +6.0 & 0 & 0.95 & 1.8$\times$ \\ 
djpeg & 1245 & 189.0 & +6.0 & +11.0 & 1.03 & 34.3$\times$ \\ 
jhead & 405 & 88.0 & 0 & 0 & 1.00 & 21.7$\times$ \\ 
libpng & 1673 & 31.0 & 0 & 0 & 1.00 & 70.9$\times$ \\ 
lodepng & 1531 & 100.3 & +6.3 & +4.7 & 0.98 & 75.6$\times$ \\ 
objdump & 992 & 146.0 & +4.0 & 0 & 0.97 & 30.6$\times$ \\ 
optipng & 1740 & 42.0 & 0 & 0 & 1.00 & 67.3$\times$ \\ 
readelf & 1055 & 150.0 & +8.0 & 0 & 0.95 & 69.5$\times$ \\ 
tcpdump & 409 & 58.3 & +9.7 & +28.7 & 1.28 & 37.3$\times$ \\ 
tiff2pdf & 3084 & 164.0 & +9.0 & 0 & 0.95 & 28.1$\times$ \\ 

\hline\hline

{\sc G. Mean} & 1335.4 & 118.7 & +5.3 & +9.1 & 1.02 & 31.2$\times$ \\ 

    \hline
  \end{tabular}
  \end{adjustbox}
\end{table}

%% file: table-lava-m.tex

\begin{table}[t]
  \vspace{-3mm}
  \centering
  \caption{Bugs found 
  on LAVA-M in 5H: avg (max) number over 5 runs.
  \label{tab:fuzzy-lava-m}} %
  \vspace{-2mm}
  \begin{adjustbox}{width=0.85\columnwidth,center}
  \begin{tabular}{ |l||c|c|c|c| }
    \hline
    &
    base64 &
    md5sum &
    uniq &
    who
    \\
    \hline
    {\sc QSYM with Z3} & 48 (48) & 58 (58) & 19 (29) &  743 (795) \\
    {\sc QSYM with Fuzzy-Sat} & 48 (48) & 61 (61) & 19.7 (29) & 2256.5 (2268) \\
    \hline 
  \end{tabular}
  \end{adjustbox}
\end{table}

%% file: table-fuzzy-stats.tex

\begin{table}[t]
  \centering
  \caption{Effectiveness of the different mutations from \fuzzysat: I2S (Input-to-State), BF (R.I. Brute Force), IC (Interesting Constants), GD (Gradient Descent), D+ND (Deterministic and Non-Deterministic mutations), MGS (Multi-Goal Strategy).\label{tab:fuzzy-mutations-stats}} %
  \vspace{-1.5mm}
  \begin{adjustbox}{width=0.88\columnwidth,center}
  \begin{tabular}{ |l||c|c|c|c|c|c| }
    \hline
    {\sc Program} 
    & {\sc I2S}
    & {\sc BF} 
    & {\sc IC}  
    & {\sc GD}
    & {\sc D+ND}
    & {\sc MGS}
    \\\hline\hline

advmng & 176 & 31 & 74 & 2 & 18 & 0 \\ 
bloaty & 43 & 5 & 20 & 5 & 19 & 4 \\ 
bsdtar & 14 & 8 & 11 & 0 & 0 & 91 \\ 
djpeg & 98 & 29 & 28 & 6 & 14 & 25 \\ 
jhead & 27 & 5 & 41 & 6 & 8 & 0 \\ 
libpng & 14 & 8 & 7 & 1 & 1 & 0 \\ 
lodepng & 61 & 6 & 16 & 1 & 21 & 0 \\ 
objdump & 91 & 21 & 18 & 1 & 10 & 5 \\ 
optipng & 28 & 7 & 6 & 0 & 1 & 0 \\ 
readelf & 96 & 10 & 22 & 0 & 22 & 0 \\ 
tcpdump & 28 & 7 & 11 & 1 & 11 & 29 \\ 
tiff2pdf & 107 & 25 & 6 & 0 & 2 & 24 \\ 

\hline\hline

{\sc Perc. on total} & 51.04\% & 10.56\% & 17.01\% & 1.50\% & 8.28\% & 11.60\% \\

    \hline
  \end{tabular}
  \end{adjustbox}
\end{table}

%% file: conclusions.tex

\edits{
\section{Threats to validity and limitations}

{\bf Floating-point arithmetic.} Our current implementation of \fuzzolic\ and \fuzzysat\ does not handle symbolic expressions involving floating-point operations. In case of floating-point instructions during program execution, \fuzzolic\ concretizes the symbolic expressions. Although this is the same strategy adopted by \qsym, we acknowledge that it may harm the effectiveness of the concolic executor on programs heavily based on floating-point computations.  

\smallskip
{\bf Order of the mutations in \bfseries{{\scshape Fuzzy-Sat}}.} The current implementation of \fuzzysat\ applies mutations using a specific order and stops as soon as one of the mutations is successful in finding a valid assignment for a query. In particular, \fuzzysat\ runs first the {\em cheapest} rules that are more {\em likely} able to succeed: e.g., given an input-to-state relation, trying the input-to-state rule first makes sense as it requires a few attempts and has high chances to succeed~\cite{redqueen}. Hence, results reported in Section~\ref{se:evaluation} are based on the order currently adopted by \fuzzolic. An interesting experiment would be to evaluate how \fuzzysat\ would perform when changing the order of the mutations.

\smallskip
{\bf Impact of \bfseries{{\scshape Fuzzy-Sat}} in hybrid fuzzing}. In Section~\ref{se:evaluation}, we have investigated the impact of \fuzzysat\ inside two concolic executors: \fuzzolic\ and \qsym. Our results are promising and suggest that \fuzzysat\ can be beneficial in the context of hybrid fuzzing. However, we believe it would be interesting to integrate \fuzzysat\ in other frameworks to have additional insights on its effect. For instance, the benefit from using \fuzzysat\ inside a concolic executor that is slow at building symbolic expression would be marginal as most of the analysis time would be spent in the emulation phase and not in the solving one. On the other hand, an efficient concolic executor should benefit from \fuzzysat\ as long as the number of queries submitted to the solver during an experiment is very high: if the number of queries is low, then a slower but more accurate traditional SMT solver would likely perform better than \fuzzysat.

}

\smallskip
\section{Conclusions}

\edits{

\fuzzysat\ is an approximate solver that uses fuzzing techniques to efficiently solve queries generated by concolic execution, helping hybrid fuzzers scale better on several real-world programs. 

We have currently identified two interesting future directions. First, we plan to integrate \fuzzysat\ in the concolic execution framework \symcc, which has been shown to be very efficient at building symbolic expressions and thus should benefit from using an efficient approximate solver. Second, we would like to devise an effective heuristic for dynamically switching during the exploration between \fuzzysat\ and a traditional SMT solver depending on the workload generated by the concolic executor on the solver backend.

}

\balance

%% file: paper.bbl
\begin{thebibliography}{10}
\providecommand{\url}[1]{#1}
\csname url@samestyle\endcsname
\providecommand{\newblock}{\relax}
\providecommand{\bibinfo}[2]{#2}
\providecommand{\BIBentrySTDinterwordspacing}{\spaceskip=0pt\relax}
\providecommand{\BIBentryALTinterwordstretchfactor}{4}
\providecommand{\BIBentryALTinterwordspacing}{\spaceskip=\fontdimen2\font plus
\BIBentryALTinterwordstretchfactor\fontdimen3\font minus
  \fontdimen4\font\relax}
\providecommand{\BIBforeignlanguage}[2]{{%
\expandafter\ifx\csname l@#1\endcsname\relax
\typeout{** WARNING: IEEEtran.bst: No hyphenation pattern has been}%
\typeout{** loaded for the language `#1'. Using the pattern for}%
\typeout{** the default language instead.}%
\else
\language=\csname l@#1\endcsname
\fi
#2}}
\providecommand{\BIBdecl}{\relax}
\BIBdecl

\bibitem{AFL}
M.~Zalewski, ``{American Fuzzy Lop},'' \url{https://github.com/Google/AFL},
  2019, [Online; accessed 20-Aug-2020].

\bibitem{AFLPP}
M.~Heuse, H.~Ei{\ss}feldt, and A.~Fioraldi, ``{AFL++},''
  \url{https://github.com/vanhauser-thc/AFLplusplus}, 2019, [Online; accessed
  20-Aug-2020].

\bibitem{Barrett2018}
C.~Barrett and C.~Tinelli, \emph{Satisfiability Modulo Theories}.\hskip 1em
  plus 0.5em minus 0.4em\relax Springer International Publishing, 2018, pp.
  305--343.

\bibitem{QSYM}
\BIBentryALTinterwordspacing
I.~Yun, S.~Lee, M.~Xu, Y.~Jang, and T.~Kim, ``{QSYM}: A practical concolic
  execution engine tailored for hybrid fuzzing,'' in \emph{Proceedings of the
  27th USENIX Conference on Security Symposium}, ser. SEC'18, 2018, pp.
  745--761. [Online]. Available:
  \url{http://dl.acm.org/citation.cfm?id=3277203.3277260}
\BIBentrySTDinterwordspacing

\bibitem{SYMCC}
\BIBentryALTinterwordspacing
S.~Poeplau and A.~Francillon, ``Symbolic execution with symcc:
  Don{\textquoteright}t interpret, compile!'' in \emph{Proceedings of the 29th
  {USENIX} Security Symposium ({USENIX} Security 20)}.\hskip 1em plus 0.5em
  minus 0.4em\relax {USENIX} Association, Aug. 2020, pp. 181--198. [Online].
  Available:
  \url{https://www.usenix.org/conference/usenixsecurity20/presentation/poeplau}
\BIBentrySTDinterwordspacing

\bibitem{stephens2016driller}
\BIBentryALTinterwordspacing
N.~Stephens, J.~Grosen, C.~Salls, A.~Dutcher, R.~Wang, J.~Corbetta,
  Y.~Shoshitaishvili, C.~Kruegel, and G.~Vigna, ``Driller: Augmenting fuzzing
  through selective symbolic execution.'' in \emph{Proceeings of the 23th
  Annual Network and Distributed System Security Symposium, {NDSS}}, 2016.
  [Online]. Available:
  \url{https://www.ndss-symposium.org/wp-content/uploads/2017/09/driller-augmenting-fuzzing-through-selective-symbolic-execution.pdf}
\BIBentrySTDinterwordspacing

\bibitem{Z3-TACS08}
\BIBentryALTinterwordspacing
L.~De~Moura and N.~Bj{\o}rner, ``Z3: An efficient smt solver,'' in
  \emph{Proceedings of 14th Int. Conf. on Tools and Algorithms for the
  Construction and Analysis of Systems}, ser. TACAS'08/ETAPS'08, 2008, pp.
  337--340. [Online]. Available:
  \url{https://doi.org/10.1007/978-3-540-78800-3_24}
\BIBentrySTDinterwordspacing

\bibitem{JFS}
\BIBentryALTinterwordspacing
D.~Liew, C.~Cadar, A.~F. Donaldson, and J.~R. Stinnett, ``Just fuzz it: Solving
  floating-point constraints using coverage-guided fuzzing,'' in
  \emph{Proceedings of the 2019 27th ACM Joint Meeting on European Software
  Engineering Conference and Symposium on the Foundations of Software
  Engineering}, ser. ESEC/FSE 2019.\hskip 1em plus 0.5em minus 0.4em\relax New
  York, NY, USA: Association for Computing Machinery, 2019, p. 521–532.
  [Online]. Available: \url{https://doi.org/10.1145/3338906.3338921}
\BIBentrySTDinterwordspacing

\bibitem{LAVAM}
\BIBentryALTinterwordspacing
B.~{Dolan-Gavitt}, P.~{Hulin}, E.~{Kirda}, T.~{Leek}, A.~{Mambretti},
  W.~{Robertson}, F.~{Ulrich}, and R.~{Whelan}, ``{LAVA: Large-Scale Automated
  Vulnerability Addition},'' in \emph{Proceedings of the 2016 IEEE Symposium on
  Security and Privacy}, ser. SP '16, 2016, pp. 110--121. [Online]. Available:
  \url{https://doi.org/10.1109/SP.2016.15}
\BIBentrySTDinterwordspacing

\bibitem{ECLIPSER-ICSE19}
\BIBentryALTinterwordspacing
J.~Choi, J.~Jang, C.~Han, and S.~K. Cha, ``Grey-box concolic testing on binary
  code,'' in \emph{Proceedings of the 41st International Conference on Software
  Engineering}, ser. ICSE '19, 2019, pp. 736--747. [Online]. Available:
  \url{https://doi.org/10.1109/ICSE.2019.00082}
\BIBentrySTDinterwordspacing

\bibitem{art-software-testing}
G.~J. Myers, C.~Sandler, and T.~Badgett, \emph{The art of software testing},
  3rd~ed.\hskip 1em plus 0.5em minus 0.4em\relax Hoboken and N.J: John Wiley \&
  Sons, 2012.

\bibitem{SurveySymbolic}
\BIBentryALTinterwordspacing
R.~Baldoni, E.~Coppa, D.~C. D'Elia, C.~Demetrescu, and I.~Finocchi, ``A survey
  of symbolic execution techniques,'' \emph{ACM Computing Surveys}, vol.~51,
  no.~3, pp. 50:1--50:39, 2018. [Online]. Available:
  \url{http://doi.acm.org/10.1145/3182657}
\BIBentrySTDinterwordspacing

\bibitem{fuzzing-book}
A.~Zeller, R.~Gopinath, M.~B\"{o}hme, G.~Fraser, and C.~Holler, ``{The Fuzzing
  Book},'' \url{https://www.fuzzingbook.org/}, 2019, [Online; accessed
  20-Aug-2020].

\bibitem{ANGR-SSP16}
\BIBentryALTinterwordspacing
Y.~Shoshitaishvili, R.~Wang, C.~Salls, N.~Stephens, M.~Polino, A.~Dutcher,
  J.~Grosen, S.~Feng, C.~Hauser, C.~Kruegel, and G.~Vigna, ``{SOK:} (state of)
  the art of war: Offensive techniques in binary analysis,'' in
  \emph{Proceedings of the 2016 {IEEE} Symposium on Security and Privacy}, ser.
  SP'16, 2016, pp. 138--157. [Online]. Available:
  \url{http://dx.doi.org/10.1109/SP.2016.17}
\BIBentrySTDinterwordspacing

\bibitem{S2E-TOCS12}
\BIBentryALTinterwordspacing
V.~Chipounov, V.~Kuznetsov, and G.~Candea, ``The {S2E} platform: Design,
  implementation, and applications,'' \emph{ACM Trans. on Computer Systems
  (TOCS)}, vol.~30, no.~1, pp. 2:1--2:49, 2012. [Online]. Available:
  \url{http://doi.acm.org/10.1145/2110356.2110358}
\BIBentrySTDinterwordspacing

\bibitem{KLEE-OSDI08}
\BIBentryALTinterwordspacing
C.~Cadar, D.~Dunbar, and D.~Engler, ``{KLEE: Unassisted and Automatic
  Generation of High-coverage Tests for Complex Systems Programs},'' in
  \emph{Proceedings of the 8th USENIX Conference on Operating Systems Design
  and Implementation}, ser. OSDI'08.\hskip 1em plus 0.5em minus 0.4em\relax
  Berkeley, CA, USA: USENIX Association, 2008, pp. 209--224. [Online].
  Available: \url{http://dl.acm.org/citation.cfm?id=1855741.1855756}
\BIBentrySTDinterwordspacing

\bibitem{SPF}
\BIBentryALTinterwordspacing
C.~S. Pasareanu, P.~C. Mehlitz, D.~H. Bushnell, K.~Gundy-Burlet, M.~Lowry,
  S.~Person, and M.~Pape, ``Combining unit-level symbolic execution and
  system-level concrete execution for testing nasa software,'' in
  \emph{Proceedings of the 2008 International Symposium on Software Testing and
  Analysis}, ser. ISSTA '08, 2008, p. 15–26. [Online]. Available:
  \url{https://doi.org/10.1145/1390630.1390635}
\BIBentrySTDinterwordspacing

\bibitem{MAYHEM}
\BIBentryALTinterwordspacing
S.~K. {Cha}, T.~{Avgerinos}, A.~{Rebert}, and D.~{Brumley}, ``Unleashing mayhem
  on binary code,'' in \emph{Proceedings of the 2012 IEEE Symposium on Security
  and Privacy}, ser. SP '12, 2012, pp. 380--394. [Online]. Available:
  \url{https://doi.org/10.1109/SP.2012.31}
\BIBentrySTDinterwordspacing

\bibitem{BCDD-CSCML17}
\BIBentryALTinterwordspacing
R.~Baldoni, E.~Coppa, D.~C. D'Elia, and C.~Demetrescu, ``{Assisting Malware
  Analysis with Symbolic Execution: A Case Study},'' in \emph{Proceedings of
  the 2017 Cyber Security Cryptography and Machine Learning}, ser. CSCML
  '17.\hskip 1em plus 0.5em minus 0.4em\relax Springer International
  Publishing, 2017, pp. 171--188. [Online]. Available:
  \url{https://doi.org/10.1007/978-3-319-60080-2_12}
\BIBentrySTDinterwordspacing

\bibitem{BCDD-CSCML19}
\BIBentryALTinterwordspacing
L.~Borzacchiello, E.~Coppa, D.~C. D'Elia, and C.~Demetrescu, ``{Reconstructing
  C2 Servers for Remote Access Trojans with Symbolic Execution},'' in
  \emph{Proceedings of the 2019 Cyber Security Cryptography and Machine
  Learning}, ser. CSCML '19.\hskip 1em plus 0.5em minus 0.4em\relax Springer
  International Publishing, 2019. [Online]. Available:
  \url{https://doi.org/10.1007/978-3-030-20951-3_12}
\BIBentrySTDinterwordspacing

\bibitem{SAGE-NDSS08}
\BIBentryALTinterwordspacing
P.~Godefroid, M.~Y. Levin, and D.~A. Molnar, ``{Automated Whitebox Fuzz
  Testing},'' in \emph{Proceedings of the 2008 Network and Distributed System
  Security Symposium}, ser. NDSS'08, 2008. [Online]. Available:
  \url{http://www.isoc.org/isoc/conferences/ndss/08/papers/10_automated_whitebox_fuzz.pdf}
\BIBentrySTDinterwordspacing

\bibitem{CDD-ASE17}
\BIBentryALTinterwordspacing
E.~Coppa, D.~C. D'Elia, and C.~Demetrescu, ``{Rethinking Pointer Reasoning in
  Symbolic Execution},'' in \emph{Proceedings of the 32nd IEEE/ACM
  International Conference on Automated Software Engineering}, ser. ASE '17,
  2017, pp. 613--618. [Online]. Available:
  \url{https://doi.org/10.1109/ASE.2017.8115671}
\BIBentrySTDinterwordspacing

\bibitem{BCE-STVR19}
\BIBentryALTinterwordspacing
L.~Borzacchiello, E.~Coppa, D.~C. D'Elia, and C.~Demetrescu, ``{Memory models
  in symbolic execution: key ideas and new thoughts},'' \emph{Software Testing,
  Verification and Reliability}, vol.~29, no.~8, 2019. [Online]. Available:
  \url{https://doi.org/10.1002/stvr.1722}
\BIBentrySTDinterwordspacing

\bibitem{PANGOLIN-SP20}
\BIBentryALTinterwordspacing
H.~Huang, P.~Yao, R.~Wu, Q.~Shi, and C.~Zhang, ``Pangolin: Incremental hybrid
  fuzzing with polyhedral path abstraction,'' in \emph{Proceedings of the 2020
  IEEE Symposium on Security and Privacy (SP)}, 2020, pp. 1613--1627. [Online].
  Available:
  \url{https://doi.ieeecomputersociety.org/10.1109/SP40000.2020.00063}
\BIBentrySTDinterwordspacing

\bibitem{TRIDENT-ISSTA20}
\BIBentryALTinterwordspacing
P.~Yao, Q.~Shi, H.~Huang, and C.~Zhang, ``Fast bit-vector satisfiability,'' in
  \emph{Proceedings of the 29th ACM SIGSOFT International Symposium on Software
  Testing and Analysis}, ser. ISSTA '20, 2020, p. 38–50. [Online]. Available:
  \url{https://doi.org/10.1145/3395363.3397378}
\BIBentrySTDinterwordspacing

\bibitem{aflsmart-tse}
\BIBentryALTinterwordspacing
V.~{Pham}, M.~{Boehme}, A.~E. {Santosa}, A.~R. {Caciulescu}, and
  A.~{Roychoudhury}, ``Smart greybox fuzzing,'' \emph{IEEE Transactions on
  Software Engineering}, 2019. [Online]. Available:
  \url{https://doi.org/10.1109/TSE.2019.2941681}
\BIBentrySTDinterwordspacing

\bibitem{google-oss-fuzz}
``{Google OSS-Fuzz: continuous fuzzing of open source software},''
  \url{https://github.com/google/oss-fuzz}, 2019, [Online; accessed
  20-Aug-2020].

\bibitem{DRILLER-NDSS16}
\BIBentryALTinterwordspacing
N.~Stephens, J.~Grosen, C.~Salls, A.~Dutcher, R.~Wang, J.~Corbetta,
  Y.~Shoshitaishvili, C.~Kruegel, and G.~Vigna, ``Driller: Augmenting fuzzing
  through selective symbolic execution,'' in \emph{Proceeings of the 23nd
  Annual Network and Distributed System Security Symposium}, ser. NDSS'16,
  2016. [Online]. Available:
  \url{http://www.internetsociety.org/sites/default/files/blogs-media/driller-augmenting-fuzzing-through-selective-symbolic-execution.pdf}
\BIBentrySTDinterwordspacing

\bibitem{lafintel}
``{Circumventing Fuzzing Roadblocks with Compiler Transformations},''
  \url{https://lafintel.wordpress.com/2016/08/15/circumventing-fuzzing-roadblocks-with-compiler-transformations/},
  2016, [Online; accessed 20-Aug-2020].

\bibitem{vuzzer}
\BIBentryALTinterwordspacing
S.~Rawat, V.~Jain, A.~Kumar, L.~Cojocar, C.~Giuffrida, and H.~Bos, ``Vuzzer:
  Application-aware evolutionary fuzzing,'' in \emph{24th Annual Network and
  Distributed System Security Symposium, {NDSS}}, 2017. [Online]. Available:
  \url{https://www.ndss-symposium.org/ndss2017/ndss-2017-programme/vuzzer-application-aware-evolutionary-fuzzing/}
\BIBentrySTDinterwordspacing

\bibitem{debray-scam14}
\BIBentryALTinterwordspacing
B.~Yadegari and S.~Debray, ``Bit-level taint analysis,'' in \emph{2014 IEEE
  14th International Working Conference on Source Code Analysis and
  Manipulation (SCAM)}, 2014, pp. 255--264. [Online]. Available:
  \url{https://doi.ieeecomputersociety.org/10.1109/SCAM.2014.43}
\BIBentrySTDinterwordspacing

\bibitem{angora}
\BIBentryALTinterwordspacing
P.~{Chen} and H.~{Chen}, ``Angora: Efficient fuzzing by principled search,'' in
  \emph{Proceedings of the 2018 IEEE Symposium on Security and Privacy (SP)},
  2018, pp. 711--725. [Online]. Available:
  \url{https://doi.org/10.1109/SP.2018.00046}
\BIBentrySTDinterwordspacing

\bibitem{redqueen}
\BIBentryALTinterwordspacing
C.~Aschermann, S.~Schumilo, T.~Blazytko, R.~Gawlik, and T.~Holz, ``{REDQUEEN:}
  fuzzing with input-to-state correspondence,'' in \emph{Proceedings of the
  26th Annual Network and Distributed System Security Symposium, {NDSS}}, 2019.
  [Online]. Available:
  \url{https://www.ndss-symposium.org/ndss-paper/redqueen-fuzzing-with-input-to-state-correspondence/}
\BIBentrySTDinterwordspacing

\bibitem{WEIZZ}
\BIBentryALTinterwordspacing
A.~Fioraldi, D.~C. D'Elia, and E.~Coppa, ``{WEIZZ: Automatic Grey-Box Fuzzing
  for Structured Binary Formats},'' in \emph{Proceedings of the 29th ACM
  SIGSOFT International Symposium on Software Testing and Analysis (ISSTA
  ’20)}, 2020. [Online]. Available:
  \url{https://doi.org/10.1145/3395363.3397372}
\BIBentrySTDinterwordspacing

\bibitem{SLF}
\BIBentryALTinterwordspacing
W.~You, X.~Liu, S.~Ma, D.~Perry, X.~Zhang, and B.~Liang, ``{SLF}: Fuzzing
  without valid seed inputs,'' in \emph{Proceedings of the 41st International
  Conference on Software Engineering}, ser. ICSE '19, 2019, pp. 712--723.
  [Online]. Available: \url{https://doi.org/10.1109/ICSE.2019.00080}
\BIBentrySTDinterwordspacing

\bibitem{wrapped-intervals}
\BIBentryALTinterwordspacing
J.~A. Navas, P.~Schachte, H.~S{\o}ndergaard, and P.~J. Stuckey,
  ``Signedness-agnostic program analysis: Precise integer bounds for low-level
  code,'' in \emph{Proceedings of the 10th Asian Symposium on Programming
  Languages and Systems}, ser. APLAS '12, 2012, pp. 115--130. [Online].
  Available: \url{https://doi.org/10.1007/978-3-642-35182-2\_9}
\BIBentrySTDinterwordspacing

\bibitem{Pin-PLDI05}
\BIBentryALTinterwordspacing
C.-K. Luk, R.~Cohn, R.~Muth, H.~Patil, A.~Klauser, G.~Lowney, S.~Wallace, V.~J.
  Reddi, and K.~Hazelwood, ``Pin: Building customized program analysis tools
  with dynamic instrumentation,'' ser. PLDI '05.\hskip 1em plus 0.5em minus
  0.4em\relax ACM, 2005. [Online]. Available:
  \url{http://doi.acm.org/10.1145/1065010.1065034}
\BIBentrySTDinterwordspacing

\bibitem{DCNPC-ASIACCS19}
\BIBentryALTinterwordspacing
D.~C. D'Elia, E.~Coppa, S.~Nicchi, F.~Palmaro, and L.~Cavallaro, ``{SoK: Using
  Dynamic Binary Instrumentation for Security (And How You May Get Caught Red
  Handed)},'' in \emph{Proceedings of the 14th ACM ASIA Conference on Computer
  and Communications Security}, ser. ASIACCS '19, 2019. [Online]. Available:
  \url{https://doi.org/10.1145/3321705.3329819}
\BIBentrySTDinterwordspacing

\bibitem{qsym-removed-fuzzbench}
Google, ``{FuzzBench: issue \#131},''
  \url{https://github.com/google/fuzzbench/issues/131}, 2020, [Online; accessed
  20-Aug-2020].

\bibitem{small-files-project}
M.~Bynens, ``{Smallest possible syntactically valid files of different
  types},'' \url{https://github.com/mathiasbynens/small}, 2019, [Online;
  accessed 20-Aug-2020].

\bibitem{jfs-custom-seeds}
D.~Liew, ``{JFS: issue \#4},''
  \url{https://github.com/mc-imperial/jfs/issues/4}, 2019, [Online; accessed
  20-Aug-2020].

\bibitem{libfuzzer}
K.~Serebryany, ``{libFuzzer: a library for coverage-guided fuzz testing},''
  \url{http://llvm.org/docs/LibFuzzer.html}, 2015, [Online; accessed
  20-Aug-2020].

\bibitem{jfs-c-interface}
R.~J. Stinnett, ``{JFS: issue \#22},''
  \url{https://github.com/mc-imperial/jfs/issues/22}, 2019, [Online; accessed
  20-Aug-2020].

\bibitem{aflpp-colorization}
``{CmpLog instrumentation for QEMU inspired by Redqueen},''
  \url{https://aflplus.plus/features/}, 2020, [Online; accessed 20-Aug-2020].

\bibitem{AFLPP-WOOT20}
\BIBentryALTinterwordspacing
A.~Fioraldi, D.~Maier, H.~Ei{\ss}feldt, and M.~Heuse, ``{AFL++: Combining
  incremental steps of fuzzing research},'' in \emph{Proceedings of the 14th
  USENIX Workshop on Offensive Technologies}, ser. WOOT '20, 2020. [Online].
  Available:
  \url{http://www.internetsociety.org/sites/default/files/blogs-media/driller-augmenting-fuzzing-through-selective-symbolic-execution.pdf}
\BIBentrySTDinterwordspacing

\bibitem{aflpp-parallel-mode}
``{Single-system parallelization},''
  \url{https://aflplus.plus/docs/parallel_fuzzing/}, 2020, [Online; accessed
  20-Aug-2020].

\bibitem{COLL-AFL}
\BIBentryALTinterwordspacing
S.~{Gan}, C.~{Zhang}, X.~{Qin}, X.~{Tu}, K.~{Li}, Z.~{Pei}, and Z.~{Chen},
  ``Collafl: Path sensitive fuzzing,'' in \emph{2018 IEEE Symposium on Security
  and Privacy}, ser. SP '18, 2018, pp. 679--696. [Online]. Available:
  \url{http://dx.doi.org/10.1109/SP.2018.00040}
\BIBentrySTDinterwordspacing

\bibitem{AFL-COV}
A.~Fioraldi, ``{afl-qemu-cov},''
  \url{https://github.com/andreafioraldi/afl-qemu-cov}, [Online; accessed
  20-Aug-2020].

\end{thebibliography}
